\documentclass[12pt]{article}

\usepackage{hyperref}
\hypersetup{
    colorlinks=true,
    linkcolor=blue,
    urlcolor=blue,
    citecolor=blue,
    linkcolor=blue
}

\usepackage{newtxtext,newtxmath}

\usepackage{graphicx}

\usepackage[letterpaper,margin=1in]{geometry}

\renewenvironment{abstract}
	{\quotation}
	{\endquotation}

\date{}

\makeatletter
\renewcommand{\fnum@figure}{\textbf{Figure \thefigure}}
\renewcommand{\fnum@table}{\textbf{Table \thetable}}
\makeatother

\usepackage{scicite}

\usepackage{url}

\def\scititle{
	Neural machine translation of seismic waves for petrophysical inversion
}
\title{\bfseries \boldmath \scititle}

\author{
	José Cunha Teixeira$^{1,2\ast}$,
        Ludovic Bodet$^{1}$,
        Agnès Rivière$^{3}$,
        Santiago G. Solazzi$^{4}$,\and
        Amélie Hallier$^{2}$,
        Alexandrine Gesret$^{3}$,
        Sanae El Janyani$^{2}$,
        Marine Dangeard$^{2}$,\and
        Amine Dhemaied$^{2}$,
        Joséphine Boisson Gaboriau$^{2}$\and
	\small$^{1}$Sorbonne Université, CNRS, EPHE, UMR 7619 METIS, 4 place Jussieu, 75252 Paris 05, France.\and
	\small$^{2}$SNCF Réseau, 6 avenue François Mitterrand, 93210 Saint-Denis, France.\and
        \small$^{3}$Geosciences Department, Mines Paris - PSL, PSL University, Paris, France.\and
        \small$^{4}$YPF Tecnología (Y-TEC) - CONICET,  B1923 Berisso, Buenos Aires, Argentina.\and
	\small$^\ast$Corresponding author. Email: jose.teixeira@sorbonne-universite.fr\and
}

\begin{document} 

\maketitle

\begin{abstract} \bfseries \boldmath

Effective structural assessment of urban infrastructure is essential for sustainable land use and resilience to climate change and natural hazards.
Seismic wave methods are widely applied in these areas for subsurface characterization and monitoring, yet they often rely on time-consuming inversion techniques that fall short in delivering comprehensive geological, hydrogeological, and geomechanical descriptions.
Here, we explore the effectiveness of a passive seismic approach coupled with artificial intelligence (AI) for monitoring geological structures and hydrogeological conditions in the context of sinkhole hazard assessment.
We introduce a deterministic petrophysical inversion technique based on a language model that decodes seismic wave velocity measurements to infer soil petrophysical and mechanical parameters as textual descriptions.
Results successfully delineate 3D subsurface structures with their respective soil nature and mechanical characteristics, while accurately predicting daily water table levels.
Validation demonstrates high accuracy, with a normalized root mean square error of 8\%, closely rivaling with conventional stochastic seismic inversion methods, while delivering broader insights into subsurface conditions 2,000 times faster.
These findings underscore the potential of advanced AI techniques to significantly enhance subsurface characterization across diverse scales, supporting decision-making for natural hazard mitigation.

\end{abstract}

\begin{figure}[p]
    \centering
    \includegraphics{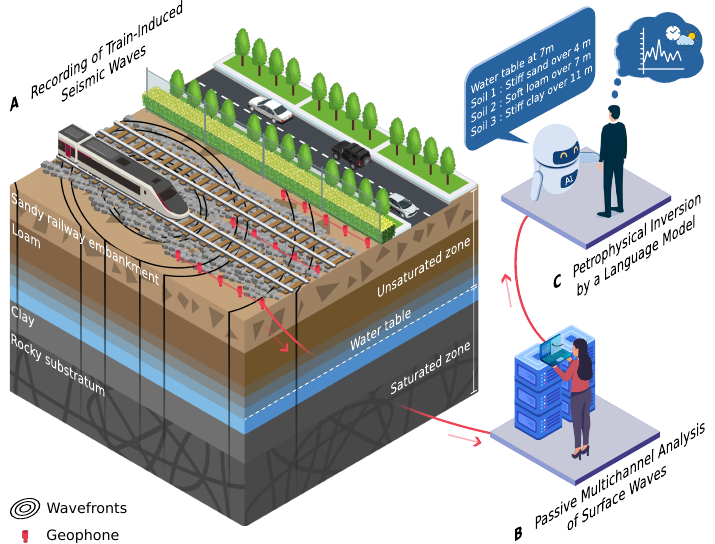}
    \caption{\textbf{Conceptual schematic description.}
    (\textbf{A}) Train-induced seismic waves are recorded along multiple geophone lines, with a focus on Rayleigh-waves.
    (\textbf{B}) These signals are analyzed through passive Multichannel Analysis of Surface Waves (passive-MASW)~\cite{Bardainne_Rondeleux_2018,Bardainne_etal_2022,Tarnus_etal_2022a,Tarnus_etal_2022b,Bardainne_etal_2023a,CunhaTeixeira_etal_2024a}, estimating a dispersion curve (Rayleigh-wave phase velocity as a function of frequency) at each geophone location along the geophone lines.
    (\textbf{C}) Each dispersion curve is subsequently inverted via a trained language model to derive a petrophysical textual description of the subsurface.
    From this descriptions, a 3D image of the subsurface can be constructed for better analysis.
    This automated procedure operates on a daily basis, allowing for continuous monitoring of subsurface hydrogeological conditions and geomechanical states over time, with respect to environmental forcing.
    }
    \label{fig:Figure1}
\end{figure}

\noindent
A shift towards low-carbon transportation systems is crucial for reducing greenhouse gas emissions and mitigating climate change.
Achieving this requires designing resilient infrastructure that can withstand evolving environmental conditions, including extreme weather and natural geohazards.
Seismic wave propagation analysis has been widely used for characterizing the Earth's subsurface~\cite{Aki_Richards_1981} and providing essential insights into geological structures.
Recent developments, leveraging continuous passive ambient seismic noise, have broadened its applications to include groundwater management and monitoring~\cite{Voisin_etal_2016,Mao_etal_2022,Shen_etal_2024,CunhaTeixeira_etal_2024b}, natural hazard assessment~\cite{Bardainne_etal_2023a,Rebert_etal_2024b}, and structural health monitoring~\cite{Bardainne_etal_2023b,CunhaTeixeira_etal_2024a}, contributing significantly to both environmental and civil engineering fields.
Some of these methods use seismic waves generated by human activity, and offer a powerful approach for characterizing the subsurface beneath critical urban infrastructure and monitoring its resilience over time (Fig.~\ref{fig:Figure1}a).

A key aspect of using seismic waves to characterize the subsurface is the need for solving inverse problems, which involves inferring subsurface properties, such as petrophysical and elastic parameters, from observed seismic data.
Whereas seismic inversion specifically focuses on estimating elastic parameters from seismic records, petrophysical inversion estimates properties related to the composition and fluid content of subsurface formations based on these elastic parameters.
Therefore, estimating petrophysical properties from seismic records involves solving two consecutive inverse problems: seismic inversion, followed by petrophysical inversion.

Inversion tasks are challenging due to the various sources of uncertainty and noise in the observed data, the nonlinearity of complex physical relationships, and the non-uniqueness of the solutions.
Thus, most approaches rely on highly regularized or stochastic methods~\cite{Tarantola_2005,Sen_Stoffa_2013,Grana_etal_2022},  which often require significant computational time and become increasingly challenging as the number of dimensions in the parameters increases.
Additionally, with the advent of advanced sensor technology and the exponential growth of the amount of data collected during seismic surveys, these methods struggle to keep up with the increasing scale and complexity of modern seismic datasets, particularly for real-time or near real-time applications.

Artificial intelligence (AI) techniques, such as deep learning~\cite{LeCun_etal_2015,Goodfellow_etal_2016}, offer promising solutions to effectively handle non-linear and complex patterns in physical problems, and have been increasingly applied for seismic and petrophysical inversion~\cite{Mousavi_Beroza_2022}.
Multilayer Perceptrons (MLPs) were used to reconstruct elastic models from raw seismic records~\cite{Araya-Polo_etal_2018}, estimate water table (WT) level maps from seismic wave velocity measurements~\cite{CunhaTeixeira_etal_2024b}, and to predict petrophysical parameters from elastic attributes~\cite{Weinzierl_Weise_2020}.
Convolutional Neural Networks (CNNs) were employed in Encoder-Decoder architectures for full-waveform inversion~\cite{Li_etal_2020,Muller_etal_2023}, petrophysical inversion~\cite{Das_Mujerki_2020}, soil type classification~\cite{Zhang_etal_2018}, and elastic model inference~\cite{Chen_etal_2022}.
Researches have also explored Recurrent Neural Networks (RNNs) for soil classification~\cite{Talarico_etal_2021}, while Graph Neural Networks (GNNs) and Generative Adversarial Networks (GANs) were suggested for inferring elastic models from seismic velocities~\cite{Liu_etal_2024} and raw seismic records~\cite{Zhang_Lin_2020}, respectively.

Nevertheless, the previous deep learning models demonstrate limited ability to capture long-range dependencies, a challenge effectively addressed by the Transformer model~\cite{Vaswani_etal_2017}, introduced in 2017, which exhibits remarkable efficiency in processing large datasets.
Transformers leverage a mechanism previously applied with RNNs, known as \textit{attention}, which assigns weights to elements within a sequence based on their relevance to each other, and allows for more accurate processing of long-range information.
This architecture has led to significant improvements in generative AI and natural language processing tasks such as neural machine translation, text generation, and speech recognition.
The well known Transformer-based GPT-3 model~\cite{Brown_etal_2020}, has demonstrated unprecedented capabilities in generating coherent and contextually relevant text, highlighting the Transformer's potential to revolutionize various applications.
The success of Transformers suggests a promising potential for generative AI in geoscience~\cite{Hadib_etal_2024}, but their application in seismic and petrophysical inversion remains relatively unexplored~\cite{Ning_etal_2023,Dou_Li_2024,Wang_etal_2023b}.

\begin{figure}[p]
    \centering
    \includegraphics[width=\textwidth]{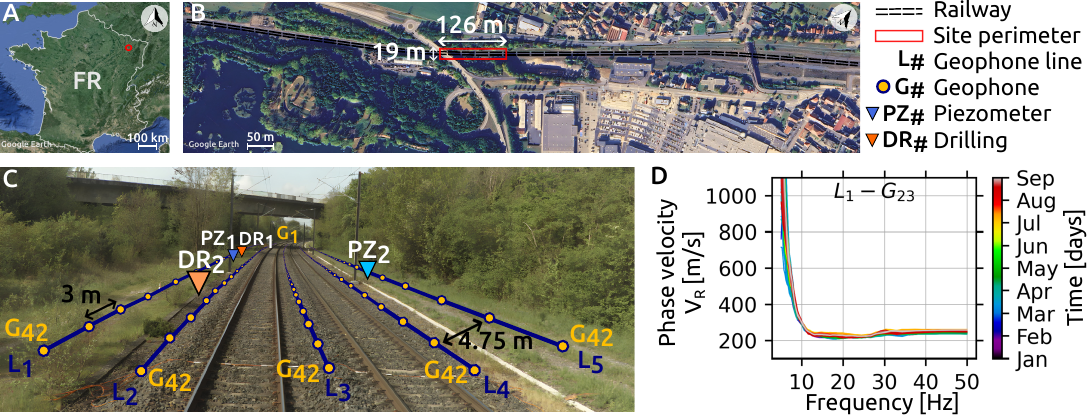}
    \caption{\textbf{Study site description.}
    (\textbf{A} and \textbf{B}) Location maps of the railway study site.
    (\textbf{C}) Experimental design of the study site showing the five geophone lines ($L_1$ to $L_5$), of 42 geophones (yellow dots $G_1$ to $G_{42}$) each, planted parallel to the railway tracks.
    Geophone lines are separated by 4.75~m, and geophones are regularly separated by 3~m.
    Rayleigh-wave dispersion curves are daily computed at each point of the geophone array by passive-MASW from seismic waves generated by train passages.
    Piezometers $PZ_1$ and $PZ_2$ were installed near geophone lines $L_1$ and $L_5$, respectively, to monitor water table levels.
    Additionally, two drillings, designated as $DR_1$ and $DR_2$, were conducted to investigate the soil formations at the site.
    (\textbf{D}) Example of Rayleigh-wave dispersion curves daily computed between January 1st, 2023, and September 31, 2023, at the position of geophone 23 of line 1 ($L_1 - G_{23}$).
    }
    \label{fig:Figure2}
\end{figure}

We leverage a challenging study site located along a railway line in the Grand-Est region of France (Fig.~\ref{fig:Figure2}a,b), at the eastern edge of the Paris Basin.
Over the last decade, this railway site has encountered several instances of sinkhole dropouts, particularly impacting the integrity of the railway.
To effectively address and mitigate the risks associated with sinkholes, continuous subsurface monitoring using passive-Multichannel Analysis of Surface Waves (passive-MASW) has been implemented since late 2020 by~\cite{Bardainne_Rondeleux_2018,Bardainne_etal_2022,Tarnus_etal_2022a,Tarnus_etal_2022b,Bardainne_etal_2023a,CunhaTeixeira_etal_2024a}.
This method uses seismic noise induced by train passages~\cite{Lavoue_etal_2020,Rebert_etal_2024a} to daily measure dispersion curves (DCs), showing Rayleigh-wave phase velocities ($V_R$) over frequencies (5 to 50~Hz), at each point of a seismic array (Figs~\ref{fig:Figure1}a,b and \ref{fig:Figure2}c,d).
Additionally, in late 2022, the site monitoring was enhanced with the addition of two piezometers, providing continuous recordings of WT levels (Fig.~\ref{fig:Figure2}c).
The substantial volume of daily available seismic data presents challenges for conventional stochastic seismic inversion methods~\cite{Sambridge_1999}, which are time-consuming and only provide a geomechanical description of the site by estimating shear-wave velocities ($V_S$) over depth from DCs.
This limitation can hinder efforts to obtain a comprehensive petrophysical and hydrogeological understanding, particularly needed when aiming to monitor both groundwater conditions and sinkhole hazards.
To address these challenges, it is essential to integrate both elastic and petrophysical properties through a rock-physics model~\cite{Mavko_etal_2009}, and to develop a frugal and fast petrophysical inversion method.

Taking inspiration from architectures used in neural machine translation and speech recognition~\cite{Baevski_etal_2020}, we explore the use of AI for deterministic petrophysical inversion by implementing a language model to generate detailed descriptions of subsurface properties based on input seismic velocity measurements (Fig.~\ref{fig:Figure1}c).
This inversion process is conceptualized as a translation task, where the seismic data is decoded into a structured textual sequence describing each soil layer in relation to the others, similar to the coherence found in language sentences.
The model, which we named \textsc{Silex} for \textit{Surface wave Inversion Lexicon}, is trained to translate DCs (numerical sequences of $V_R$) into detailed textual descriptions of the subsurface’s petrophysical parameters up to 20~m depth.
The training dataset consists of pairs of soil petrophysical descriptions and their corresponding DCs, modeled using the rock-physics model developed by~\cite{Solazzi_etal_2021}, that we adapted for multi-layer parametrization (see Materials and Methods).

This inversion approach not only enhances the efficiency and accuracy of the inversion task but also provides unique insights into subsurface petrophysical characteristics that were previously unattainable using conventional methods due to the high number of parameters to invert and the time constraint of the study.
The resulting inverted petrophysical parameters can ultimately be used for site geological and hydrogeological characterization~\cite{Shen_etal_2024}, or can be converted into effective elastic properties for structural health assessment and monitoring.

\section*{Results}
\label{sec:Results}

\subsection*{Petrophysical and mechanical properties}
\label{sec:PetrophysicalAndMechanicalProperties}
\textsc{Silex} was used to invert all subsequent DCs along the five geophone lines ($L_1$ to $L_5$), on a daily basis from September 3, 2020, to September, 4, 2023.
Each DC is translated into a textual representation that includes the WT level ranging from 0.5 to 19.5~m (step of 0.5~m), and one to four soil layers.
Each layer is characterized by a soil type (sand, loam, silt, or clay), a thickness ranging from 1 to 20~m (step of 1~m and a thicknesses sum for all layers of 20~m), and an average number of contacts per particle ($N$) ranging from 6 to 10 (step of 1) which is linked to the soil compaction level.
All four soil types are defined with ratios of pure sand ($\gamma_{sand}$), clay ($\gamma_{clay}$), and silt ($\gamma_{silt}$), porosity ($\phi$), van Genuchten parameters ($\alpha_{vg}$, $n_{vg}$, and $\theta_{vg}$)~\cite{vanGenuchten_1980}, a residual water saturation ($S_{wr}$), and a fraction of non-slipping grains ($f_{nsg}$)~\cite{Carsel_Parrish_1988,Solazzi_etal_2021} (see Materials and Methods).
Since each DC depends on the soil properties beneath the surface point where it was measured (see Materials and Methods), the resulting petrophysical parameters are concatenated along each geophone line to build 2D~sections that facilitate the visualization of complex subsurface structures and enhance the spatial understanding of geological formations and hydrogeological conditions.

The resulting petrophysical sections, monthly-averaged over July 2022 and 2023, (Fig.~\ref{fig:Figure3}, or fig.~S15 of Supplementary Text for a more detailed view), highlight a predominantly soft clayey soil ($N$ at 6 and 7), as expected from geological drillings ($DR_1$ and $DR_2$ in Figs~\ref{fig:Figure2} and \ref{fig:Figure3}).
Results also indicate the presence of compacted soil structures in the North-East, and to a lesser extent in the South-West regions of all geophone lines.
These are located at depths ranging from -5 to -10~m and are predicted to be composed of a compact sandy soil ($N$ value of 9 and 10).
At both drilling locations, \textsc{Silex} predicts a loam layer with a moderate $N$ value of 7 where the drillings reveal a sandy clay soil, and a sand layer with a high $N$ value of 10 at the depth corresponding to a gravelly sand layer.

However, the gypsum substratum remains undetected by our model as it does not qualify as soil.
Additionally, we exclusively use dispersion data above 15~Hz, which likely limits the investigation resolution at the depth of the gypsum layer.
Typically, the maximum sensitivity depth is approximately half of the maximum wavelength $\lambda_{max} = \max \left[ \frac{V_{R}(f)}{f} \right]$, where $V_R$ is the Rayleigh-wave phase velocity and $f$ is the frequency \cite{Foti_etal_2018}.
In our case, the average $\lambda_{max}$ is 24~m, corresponding to a maximum investigation depth with good resolution of approximately -12~m.
Nevertheless, a highly compacted sandy structure below -15~m was predicted around $x = 42$ m.
This prediction may approximate the mechanical behavior of the substratum layer but no drilling was conducted at this location to confirm the findings.

Drained shear modulus of the porous medium ($\mu_m$) sections were computed from the petrophysical inversion results, using a modified Hertz-Mindlin model~\cite{Mindlin_1949} (Fig.~\ref{fig:Figure4}, or fig.~S14 of Supplementary Text for a more detailed view).
Visualizing this parameter makes different structures, with the lowest and highest $\mu_m$-values, more easily discernible.

\subsection*{Water table level}
\label{sec:WaterTableLevel}
The WT levels were inferred with a depth step of 0.5~m.
Thus, to obtain a less noisy WT levels across the study area and period, a Savitzky-Golay filter~\cite{Savitzky_Golay_1964} was applied, using a spatial window of 10.75~m along the x-axis and a temporal window of 137~days spanning (September 2020 to September 2023).
Nevertheless, the depth resolution could be enhanced by reducing the depth step during training data generation and accordingly adapting the model vocabulary (see Materials and Methods).

The inferred levels align closely with the observed data at piezometers $PZ_1$ and $PZ_2$ (Fig.~\ref{fig:Figure5}), demonstrating \textsc{Silex}'s accuracy in capturing WT levels across time.
However, sharp variations in the WT are not fully recovered, and a slight overestimation of approximately 0.5~m is observed in the inferred levels at $PZ_2$.
This could be explained by the fact that $PZ_2$ is located over a relatively thick and laterally extensive gravely sand bank (Fig.~\ref{fig:Figure3}, or fig.~S15 of Supplementary Text for a more detailed view).
Since, gravely sand is not particularly sensitive to capillary suction, the corresponding mechanical properties are not expected to be particularly sensitive to saturation changes.
These results are consistent with previous studies that used an MLP to directly convert DCs into WT levels, based on observed data~\cite{CunhaTeixeira_etal_2024b}.

Additionally, the model accurately captures seasonal variations, with WT levels rising and falling in sync with low and high water periods.
These fluctuations correspond closely to regional rainfall patterns, indicating the model's sensitivity to hydrological drivers.

\subsection*{Accuracy and error}
\label{sec:AccurayAndError}
\textsc{Silex} was evaluated on 50,000~synthetic samples outside the training dataset, achieving an accuracy on the inferred tokens of 81\%.
Additionally, using \textsc{Silex} inference results, DCs were recomputed with the same rock-physics model used to generate the training data (see Materials and Methods).
The recomputed DCs were compared to the original ones used as input for the inversion, allowing the error assessment of the model’s predictions.
The average root mean square error (RMSE) on the DCs was estimated at 12~m/s (normalized RMSE of 8\%), across all inversions from September 4, 2020, to September 3, 2023, which is impressively low (see Materials and Methods).

\section*{Discussion}
\label{sec:Discussion}

\subsection*{Temporal stability}
\label{sec:TemporalStability}
Subsurface structures above -15~m demonstrate consistent temporal stability over the study period, in terms of both inferred soil types and $\mu_m$.
This stability is crucial for understanding the long-term geological and hydrogeological conditions of the area, as well as for sinkhole risk assessment.
However, below -15~m, the substratum imaging exhibits less stability (substratum at the North-East part of $L_1$ on Fig.~\ref{fig:Figure4}b and d).
This can be attributed to the frequency bandwidth limitations of the inverted DCs (15 to 50~Hz) rather to \textsc{Silex}'s inference capabilities, as the frequency lower limit may not sufficiently resolve the deeper substratum features, leading to a less stable and accurate representation of the subsurface characteristics below -15~m.

The observed evolution of $\mu_m$ does not show a significant development of mechanical weaker zones (low $\mu_m$).
This is coherent with the fact that no evidence of sinkhole development was observed at the surface of the site during the study period.
Nevertheless, the uncertainty and reduced stability of the substratum at greater depths necessitate careful consideration.

\subsection*{Comparison with conventional inversion}
\label{sec:ComparisonWithConventionalSeismicInversion}
The $V_S$ estimations, derived from \textsc{Silex}'s inference results (see Materials and Methods), closely align with those obtained with a conventional seismic inversion stochastic method based on a \textit{Neighborhood Algorithm}~\cite{Sambridge_1999,Wathelet_2008,Pasquet_Bodet_2017}.
This confirms that the model reliably retrieves mechanical results that are comparable to established methods, despite the radical differences between the two approaches (see Supplementary Text).

Beyond the additional detailed insights gained from a petrophysical inversion compared to a seismic inversion, \textsc{Silex} also provides a significant computational advantage.
On a single Intel Xeon central processing unit (CPU), the inversion of an entire daily geophone line is completed within 2 minutes, whereas the conventional seismic inversion requires approximately 65 hours, which is excessively time-consuming for daily monitoring.
However, it is important to note that this extended time is primarily due to the stochastic nature of the conventional approach, which necessitates computing a large number of models when sampling the posterior probability distribution.
Even though conventional seismic inversion can be optimized through parallel computing to reduce processing time, a conventional petrophysical inversion method of the rock-physics model we employed~\cite{Solazzi_etal_2021}, yet to be fully developed, would still require significantly more time due to the high number of petrophysical parameters to retrieve.

The training data generation phase is highly time-consuming, requiring up to 20 hours for 1,316,446 samples on a single Intel Xeon CPU.
However, it can be parallelized and is only necessary once per lifetime, as the data can be reused to train other models by selecting the desired frequencies.
Similarly, the model's training phase is computationally intensive, requiring around 2 hours on a Nvidia Tesla T4 GPU, but it only needs to be performed once before deployment.
Following this training, the model can be deployed for daily inversion tasks without any additional overhead, thereby demonstrating high efficiency for large-scale geophysical surveys.
The lightweight architecture of \textsc{Silex} (only 499,266 parameters, see Materials and Methods) allows it to operate on standard hardware without requiring server infrastructure, thereby enhancing its practicality for field applications and effectively addressing technical limitations in geophysical monitoring.

\subsection*{Rock-physics model limitation}
\label{sec:RockPhysicsModelLimitation}
The modified rock-physics model~\cite{Solazzi_etal_2021} used to generate the training data is primarily designed to model soil properties rather than consolidated rock formations.
As a result, the inversion is accurate at shallow depths, where soils predominate, but may not accurately reflect the petrophysical properties of the rocky substratum, even though the mechanical properties of rocks are approximated with a high $N$ (9 and 10 on Fig.~\ref{fig:Figure3}b,d).
This limitation can lead to a lack of detailed information and perceived instability at greater depths, as certain geological features and variations may be overlooked or inadequately captured.

Incorporating rock lithofacies could mitigate these issues but would require a redefinition of the rock-physics model.
This is relatively straightforward for sandstones or sedimentary rocks, but implementing the full diversity of rock types (igneous, sedimentary, and metamorphic) would necessitate a more complex rock-physics model.

\subsection*{Lack of uncertainty estimation}
\label{sec:LackOfUncertainty}
One limitation of a deterministic petrophysical inversion approach is the lack of inherent uncertainty quantification, which contrasts with stochastic methods commonly employed in seismic inversion.
In fact, the Transformer-based language model used here provides a unique output per inversion, without representing the range of possible outcomes or associated uncertainties.
While it is optimized for accuracy and efficiency, the absence of uncertainty quantification may limit its application in scenarios where risk assessment is critical.

Future enhancements could integrate uncertainty estimation methods, such as ensemble modeling, or Bayesian inversion through posterior sampling with diffusion models~\cite{Moufad_etal_2024}, to enhance the model's value in high-stakes geological and hydrogeological monitoring contexts.

\section*{Conclusions}
\label{sec:Conclusions}
In this study, we introduced a new approach for subsurface petrophysical inversion integrating a Transformer-based language model.
This model effectively translates DCs derived from train-induced surface waves into detailed textual descriptions of subsurface properties, including soil types, mechanical properties, and WT levels.
These descriptions can be used for geological and hydrogeological interpretation or converted into soil elastic moduli for applications in civil engineering.
The model effectively captured the spatial and temporal stability of subsurface structures above -15~m, correlating well with expected lithofacies and indicating the absence of sinkhole emergence at the surface during the study period.
The inferred WT levels also closely match the measured data from piezometers, and accurately reflects seasonal variations induced by rainfall.
Additionally, this technique demonstrated significant improvements in inversion speed, achieving a 2,000-fold increase compared to conventional stochastic methods.

Despite these successes, challenges remain in accurately imaging deeper substratum features below -15~m, largely due to limitations in the seismic data frequency bandwidth and the rock-physics model's design, used to generate the training data.
Future research should focus on expanding the inverted frequency range, and refining the rock-physics model to also account for rock properties at greater depths.

The lack of uncertainty estimation inherent to our deterministic approach limits the reliability of predictions for critical decision-making.
Addressing this limitation in future work, potentially through the integration of uncertainty quantification techniques would enhance the model’s trustworthiness, particularly for deeper and more complex geological features.

Overall, these findings underscore the potential of AI techniques in improving subsurface characterization and providing valuable insights for applications such as structural health assessment, natural hazard monitoring, and groundwater management.
This approach represents a promising step forward in petrophysical inversion, offering a framework for future studies aiming to enhance subsurface characterization through advanced AI methodologies across various scales.


\newpage

\newpage

\newpage
\begin{figure}[p]
    \centering
    \includegraphics[width=\textwidth]{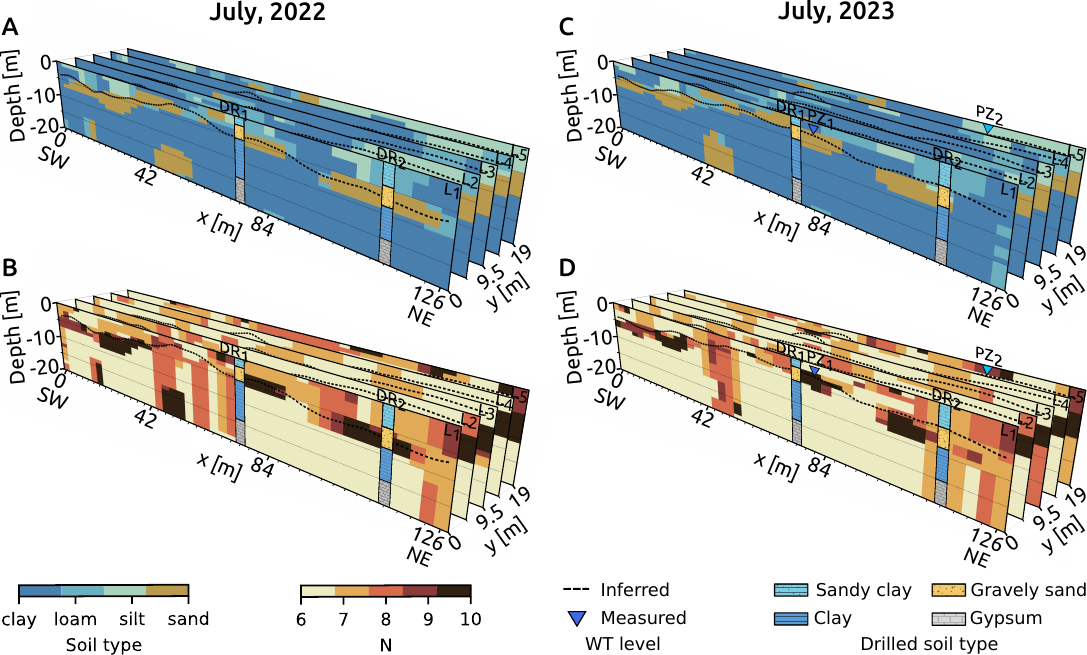}
    \caption{\textbf{Petrophysical inversion results.}
    Section representation of the petrophysical inversion results using \textsc{Silex} model, across the five geophone lines ($L_1$ to $L_5$), on July 2022 and 2023.
    (\textbf{A} and \textbf{C}) Sections of soil types and smoothed water table (WT) level (dashed line).
    (\textbf{B} and \textbf{D}) Sections of the average number of contacts per particles ($N$) and smoothed water table level (dashed line).
    The soil types identified from drillings at $DR_1$ and $DR_2$, as well as the measured water table levels at piezometers $PZ_1$ and $PZ_2$ (only in 2023), are shown for reference.
    }
    \label{fig:Figure3}
\end{figure}

\newpage
\begin{figure}[p]
    \centering
    \includegraphics[width=\textwidth]{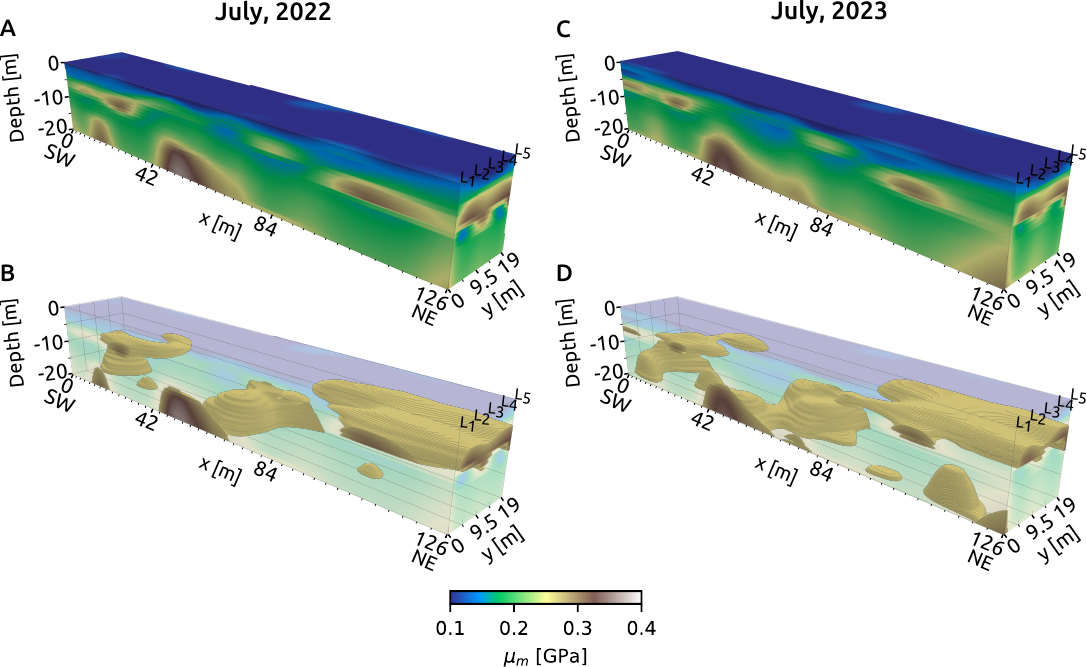}
    \caption{\textbf{Interpolated shear modulus volume.}
    Shear modulus ($\mu_m$) across the five geophone lines ($L_1$ to $L_5$) computed from the petrophysical inversion results using a modified Hertz-Mindlin model~\cite{Mindlin_1949}, on July 2022 and 2023.
    (\textbf{A} and \textbf{C}) Complete shear modulus cube.
    (\textbf{B} and \textbf{D}) Formations with $\mu_m>$ 0.25~GPa).
    }
    \label{fig:Figure4}
\end{figure}

\newpage
\begin{figure}[p]
    \centering
    \includegraphics{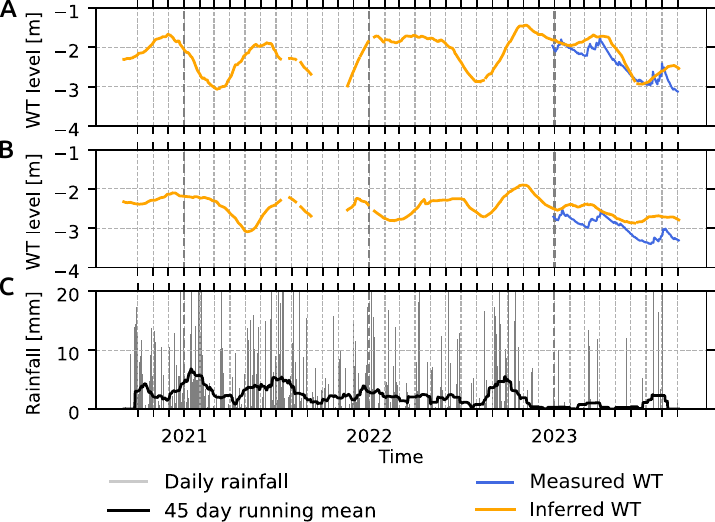}
    \caption{\textbf{Measured and inferred water table level, and rainfall over time.}
    (\textbf{A} and \textbf{B}) Comparison between measured and inferred water table level variation over time at the positions of piezometers $PZ_1$ and $PZ_2$.
    (\textbf{C}) Daily measured rainfall, and 45-day running mean.
    }
    \label{fig:Figure5}
\end{figure}


\clearpage 

%
\bibliography{science_template} 
\bibliographystyle{sciencemag}


\clearpage
\section*{Acknowledgments}
We thank Thomas Bardainne, Renaud Tarnus, Thibaut Allemand, Ceifang Cai, Helene Toubiana Lille, Nicolas Deladerriere, Lilas Vivin, and Loic Michel from the company Sercel for their contribution in acquiring and processing the geophysical passive-MASW data.
We also express gratitude to the local teams from SNCF Réseau their invaluable assistance and support during field experiments.
Finally, we thank Damien Jugnot from UMR 7619 METIS, Sorbonne Université, for the helpful discussions on rock-physics theory.

\paragraph*{Funding:}
This research was made possible through funding from SNCF Réseau, CNRS, Sorbonne Université, Mines Paris-PSL research contract, and ANRT/Cifre-SNCF Réseau n°2021/1552 convention.

\paragraph*{Author contributions:}
J.C.T. conceived the study and the language model.
L.B., A.G. A.D., and J.B.G. oversaw the study.
A.H., A.R., M.D., and S.E.J. planned and oversaw the installation of the piezometers.
A.R., and S.E.J. performed piezometric data interpretation.
J.C.T., and S.S. performed the multi-layer adaptation of the rock-physics model.
J.C.T., L.B., A.R., S.S., A.G., J.B.G., A.H., S.E.J., and M.D. contributed to the preparation of the manuscript and were involved in the interpretation of results.

\paragraph*{Competing interests:}
The processing technique of the geophysical passive-MASW data used as input to our model, was performed by the company Sercel, and is under the patents No.~FR3084473B1, BR112021000806B1, EP3827288B1, WO2020021177A1, CA3106292A1, US11971514B2, AU2019311797B2, and ZA202101057B, entitled "Method and device for monitoring the subsoil of the earth under a target zone".
The authors declare no competing interests.

\paragraph*{Data and materials availability:}
All data and code that support the claims in this manuscript are provided on Zenodo~\cite{Cunha_Teixeira_2024}.


\subsection*{Supplementary materials}
Materials and Methods\\
Supplementary Text\\
Figs S1 to S28\\
Tables S1 to S7\\
References \textit{(49-\arabic{enumiv})}\\ 


\newpage


\renewcommand{\thefigure}{S\arabic{figure}}
\renewcommand{\thetable}{S\arabic{table}}
\renewcommand{\theequation}{S\arabic{equation}}
\renewcommand{\thepage}{S\arabic{page}}
\setcounter{figure}{0}
\setcounter{table}{0}
\setcounter{equation}{0}
\setcounter{page}{1} 


\begin{center}
\section*{Supplementary Materials for\\ \scititle}

José Cunha Teixeira$^{1,2\ast}$,
Ludovic Bodet$^{1}$,
Agnès Rivière$^{3}$,
Santiago G. Solazzi$^{4}$, \\
Amélie Hallier$^{2}$,
Alexandrine Gesret$^{3}$,
Sanae El Janyani$^{2}$,
Marine Dangeard$^{2}$, \\
Amine Dhemaied$^{2}$,
Joséphine Boisson Gaboriau$^{2}$ \\
\small$^{1}$Sorbonne Université, CNRS, EPHE, UMR 7619 METIS, 4 place Jussieu, 75252 Paris 05, France. \\
\small$^{2}$SNCF Réseau, 6 avenue François Mitterrand, 93210 Saint-Denis, France. \\
\small$^{3}$Geosciences Department, Mines Paris - PSL, PSL University, Paris, France. \\
\small$^{4}$YPF Tecnología (Y-TEC) - CONICET,  B1923 Berisso, Buenos Aires, Argentina. \\
\small$^\ast$Corresponding author. Email: jose.teixeira@sorbonne-universite.fr \\
\end{center}

\subsubsection*{This PDF file includes:}
Materials and Methods\\
Supplementary Text\\
Figures S1 to S28\\
Tables S1 to S7\\


\newpage


\subsection*{Materials and Methods}
\label{sec:MaterialsAndMethods}

\subsubsection*{Passive-Multichannel Analysis of Surface Waves}
\label{sec:passive-masw}
The seismic dispersion data was acquired and processed by~\cite{Bardainne_Rondeleux_2018}.
Daily dispersion curves (DCs) were constructed using the passive-Multichannel Analysis of Surface Waves (passive-MASW) approach~\cite{Park_Miller_2008,Quiros_etal_2016,Cheng_etal_2015,Cheng_etal_2016,Bardainne_etal_2022,Mi_etal_2022,Czarny_etal_2023,Rezaeifar_etal_2023,You_etal_2023,CunhaTeixeira_etal_2024a}, which combines seismic interferometry~\cite{Bensen_etal_2007} with traditional 2D~MASW~\cite{Bohlen_etal_2004,Socco_Strobbia_2004,Socco_etal_2010,Bergamo_etal_2012,Pasquet_Bodet_2017}.

Since September 2020, seismic noise induced by train passages, has been continuously recorded using five uniform linear arrays of vertical component geophones ($L_1$ to $L_5$ on Fig.~2c of the main text).
Each linear array has a length of 123~m and is equipped with 42~3-meter spaced geophones.
The geophones were strategically positioned along the rail track, either on the cess (i.e., the track side) for linear arrays $L_1$ and $L_5$, or on the ballast for $L_2$, $L_3$, and $L_4$.
Note that our experience has shown that the ballast does not significantly affect the quality of the data.

For each day, passage events were automatically detected by taking advantage of their high signal-to-noise ratio.
Then, based on the assumption that trains act as permanent and well-localized moving sources~\cite{Bardainne_etal_2022,Rebert_etal_2023,Rebert_etal_2024a}, specific time intervals corresponding to moments when the train is outside the seismic array but within a range and angle enabling the induced waves to propagate in alignment with all the geophones along the geophone lines.
Various techniques exist for this automatic segment selection and include Frequency-Wavenumber (FK)-based data selection~\cite{Cheng_etal_2018,CunhaTeixeira_etal_2024a}, data selection in the Tau-p domain~\cite{Cheng_etal_2019}, beamforming techniques~\cite{Ning_etal_2022}, and covariance matrix methods for train localization in time and space~\cite{Rezaeifar_etal_2023}.
However, the specific selection method used by~\cite{Bardainne_Rondeleux_2018} is not known to the authors.

The selected time intervals were then divided into overlapping short segments of a few seconds each (usually around 10~s), and cross-correlations were applied on every geophone pair.
This process forms a virtual shot-gather for each segment, illustrating the propagation of seismic waves from the first sensor (acting as the virtual source) to the subsequent sensors.
To improve the reliability of the data, virtual-shot gathers resulting from all segments of multiple trains passages (approximately 40~per~day) were stacked together in the time-domain.
This stacking process enhances the signal-to-noise ratio, as random noise tends to cancel out while the consistent seismic signals reinforce each other.
Studies indicate that even a relatively small number of train passages can provide sufficient data for effective analysis~\cite{CunhaTeixeira_etal_2024a}.

Finally, a MASW window~\cite{Pasquet_Bodet_2017} of 21~m (over 8~geophones) with a moving step of 3~m (1 geophone interval) was used to extract Rayleigh-wave DCs from the stacked virtual shot-gathers along each geophone line.
Within the sliding MASW window shot-gathers were transformed from the distance-time~$(x,t)$ domain into dispersion images in the frequency-phase velocity~$(f,V_R)$ domain using a phase-shift transform~\cite{Park_etal_1999}, and the fundamental propagation mode DC was automatically picked between 5 and 50~Hz~\cite{Tarnus_etal_2022b}.
Each DC was then positioned at its associated MASW window center to construct the five $V_R$ profiles between 10.5 and 115.5~m.

In parallel, Spectral Analysis of Surface Waves or SASW~\cite{Nazarian_etal_1983} was also used to construct horizontal maps of $V_R$, for frequencies ranging from 5 to 50~Hz, between pairs of geophones.
The resulting velocities were combined with the velocities estimation from passive-MASW to construct a complete $V_R$ cube between 0 and 126~m.
Finally, at each geophone line $y$-position, DCs were interpolated and extracted from the data cube at each geophone location, and at the limits of the array.
This results in a rectangular configuration of 43~3-meter spaced data points per geophone line, as they are alternatively shifted by one geophone.

\subsubsection*{Model design}
\label{sec:ModelDesign}
\begin{figure}[p]
    \centering
    \includegraphics[width=\textwidth]{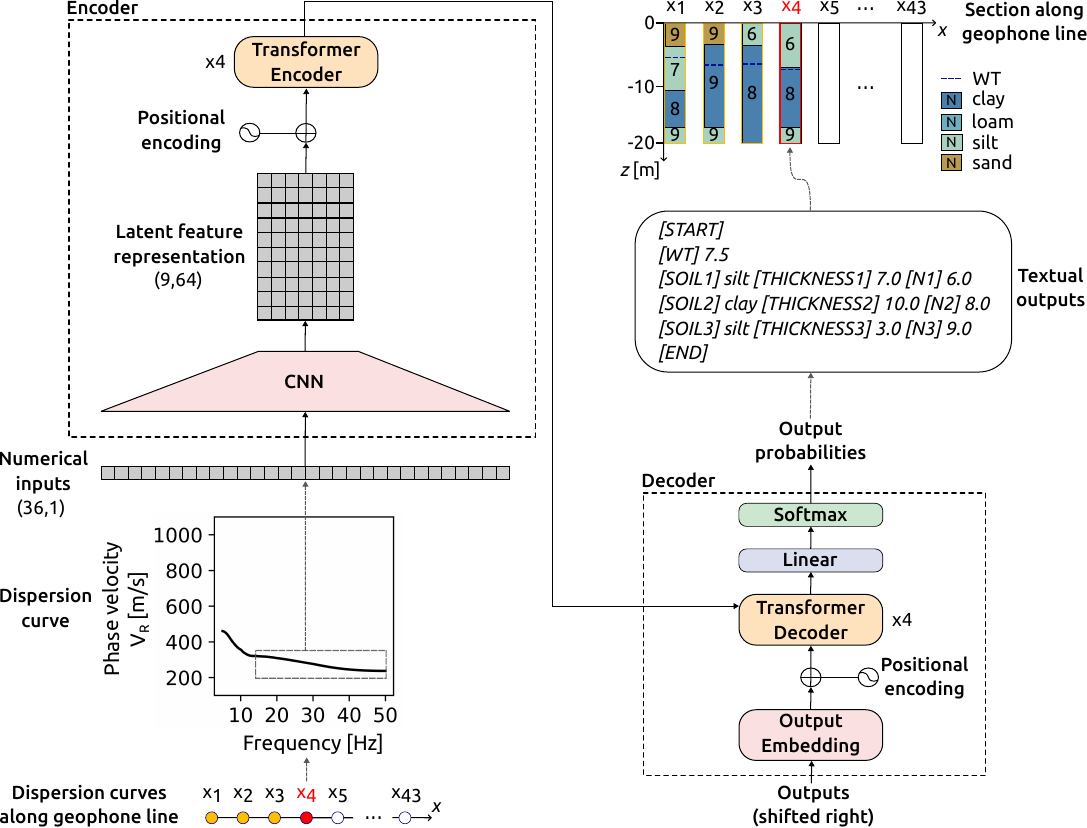}
    \caption{\textbf{\textsc{Silex} model architecture used to generate a textual sequence petrophysically describing the subsoil from a dispersion curve.}
    This architecture is inspired by the original Transformer Encoder-Decoder model~\cite{Vaswani_etal_2017}, but is adapted for numerical inputs instead of textual ones.
    The model takes as input a numerical vector of 36 values corresponding to the Rayleigh-wave phase velocities ($V_R$) at frequencies between 15 and 50~Hz, with a step of~1 Hz.
    Latent features are extracted from the input vector using a convolutional neural network (CNN) that combines convolution and max pooling operations, replacing the embedding operation used for textual inputs in the original model.
    A positional encoding is added to the latent feature tensor, which is then passed through four consecutive Transformer encoders.
    The encoder outputs are processed through four Transformer decoders, trained with teacher forcing to generate textual output based on the numerical inputs.
    The textual outputs display the water table (WT) level, and one to four soil layers, each characterized with a soil type (sand, loam, silt, clay) (table~\ref{tab:TableS1}), a thickness ranging from 1 to 20~m, with a step of 0.5~ (total thicknesses sum must be 20~m), and an average number of contacts per particle ($N$) ranging from 6 to 10.
    This textual output can be converted into a vertical log representation to be concatenated with subsequent inversions, forming a subsoil section image (Fig.~3 of the main text).
    }
    \label{fig:FigureS1}
\end{figure}

We adapted the original Transformer Encoder-Decoder architecture~\cite{Vaswani_etal_2017} to handle numerical input sequences, rather than textual data, allowing it to process Rayleigh-wave phase velocities across frequencies (15 to 50~Hz with a step of 1~Hz) as input (fig.~\ref{fig:FigureS1}).
To achieve this, we replaced the standard input embedding layer of the encoder part with a Convolutional Neural Network (CNN) that transforms the input sequence into a latent feature representation~\cite{Baevski_etal_2020}.
The CNN module comprises a sequence of three convolutional blocks with a gradual increase in the number of filters, enabling the model to learn increasingly complex features at each stage.
The first block includes two convolutional layers, each with 16 filters and a kernel size of 3, followed by a max pooling operation with a window size of 2.
The second block contains two convolutional layers with an increased filter size of 32 and a kernel size of 3, also followed by max pooling with a window size of 2.
The final block comprises two convolutional layers with 64 filters and a kernel size of 3.
A sinusoidal positional encoding layer~\cite{Vaswani_etal_2017} is then applied to the latent space, allowing the model to incorporate sequence order information into the numerical data.
This helps the model effectively learn patterns and dependencies across the input frequencies, essential for translating DCs into petrophysical descriptions.
The encoded representation is then processed through a stack of four Transformer encoder layers, each with eight attention heads and a feedforward neural network with an intermediate dimensionality of 256.
The multi-headed attention mechanism allows the model to attend to multiple aspects of the input sequence in parallel, enhancing its ability to capture complex relationships across various frequency components.

The decoder part mirrors the encoder’s architecture and is designed to generate structured petrophysical textual descriptions based on the encoded representation of the input sequence.
It follows the same original architecture~\cite{Vaswani_etal_2017}, and consists in an embedding layer of the partial generated output sequence, followed by a positional encoding layer. 
Like the encoder part, the decoder part stacks four transformer decoder layers, each with eight attention heads and a Multilayer Perceptron (MLP) with an intermediate dimensionality of 256.
However, it incorporates two distinct attention mechanisms.
A masked self-attention layer processes the partially generated output sequence, allowing each position in the sequence to attend only to earlier positions, and a cross-attention mechanism, which attends to the encoder’s final output and allows the decoder to focus on relevant aspects of the encoded DCs, ensuring that each generated token reflects the features captured from the input sequence.
Finally, each transformer decoder layer includes a feedforward network with an intermediate dimension of 256, which further refines the sequence.

The decoder's outputs a probability distribution over possible word tokens at each step, representing the likelihood of each petrophysical attribute (table~\ref{tab:TableS1}).
This distribution enables the model to select the most probable token for each position in the sequence, forming a structured textual sequence of a petrophysical description that aligns with the input DC's characteristics (see textual output on fig.~\ref{fig:FigureS1}).
This output sequence is variable in length, and is dynamically adjusted to represent the inferred number of soil layers (1 to 4).
The structured petrophysical description includes details such as soil type, layer thickness, particle contact number ($N$), and water table (WT) level, aligned with the characteristics of the input DC.
A restrictive sampler is employed at each generation step to maintain a coherent structure in the output sequence by selectively allowing or prohibiting certain tokens (table~\ref{tab:TableS2}).

\begin{table}[!htbp]
\centering
  \caption{\textbf{Petrophysical vocabulary.}
  List of tokens and corresponding index used to generate textual petrophysical descriptions.
  }
  \begin{tabular}{p{0.2\textwidth}p{0.1\textwidth}||p{0.2\textwidth}p{0.1\textwidth}}
      \hline
      Token & Index & Token & Index \\
      \hline
      $[$PAD$]$ & 0 & 5.0 & 25 \\
      $[$START$]$ & 1 & 5.5 & 26 \\
      $[$END$]$ & 2 & 6.0 & 27 \\
      $[$WT$]$ & 3 & 6.5 & 28 \\
      $[$SOIL1$]$ & 4 & 7.0 & 29 \\
      $[$SOIL2$]$ & 5 & 7.5 & 30 \\
      $[$SOIL3$]$ & 6 & 8.0 & 31 \\
      $[$SOIL4$]$ & 7 & 8.5 & 32 \\
      $[$THICKNESS1$]$ & 8 & 9.0 & 33 \\
      $[$THICKNESS2$]$ & 9 & 9.5 & 34 \\
      $[$THICKNESS3$]$ & 10 & 10.0 & 35 \\
      $[$THICKNESS4$]$ & 11 & 11.0 & 36 \\
      $[$N1$]$ & 12 & 12.0 & 37 \\
      $[$N2$]$ & 13 & 13.0 & 38 \\
      $[$N3$]$ & 14 &  14.0 & 39 \\
      $[$N4$]$ & 15 & 15.0 & 40 \\
      0.5 & 16 & 16.0 & 41 \\
      1.0 & 17 & 17.0 & 42 \\
      1.5 & 18 & 18.0 & 43 \\
      2.0 & 19 & 19.0 & 44 \\
      2.5 & 20 & 20.0 & 45 \\
      3.0 & 21 & clay & 46 \\
      3.5 & 22 & loam & 47 \\
      4.0 & 23 & silt & 48 \\
      4.5 & 24 & sand & 49 \\
      \hline
    \end{tabular}
    \label{tab:TableS1}
\end{table}

\begin{table}[!htbp]
\centering
  \caption{\textbf{Allowed tokens.}
  List of the allowed tokens at each inference step. Note that the model will pad the sequence after the $[$END$]$ token until the maximum sequence length of 24 tokens is reached.
  }
  \renewcommand{\arraystretch}{0.85}
  \begin{tabular}{l|c}
      \hline
      Step & Allowed tokens \\ [0.5ex]
      \hline\hline
      
      0 & $[$START$]$ \\
      \hline
      
      1 & $[$WT$]$ \\
      2 & 0.5, 1.0, 1.5, 2.0, 2.5, 3.0, 3.5, 4.0, 4.5, 5.0, \\
        & 5.5, 6.0, 6.5, 7.0, 7.5, 8.0, 8.5, 9.0, 9.5, 10.0 \\
      \hline
      
      3 & $[$SOIL1$]$ \\
      4 & clay, loam, silt, sand \\
      5 & $[$THICKNESS1$]$ \\
      6 & 1.0, 2.0, 3.0, 4.0, 5.0, 6.0, 7.0, 8.0, 9.0, 10.0, \\
        & 11.0, 12.0, 13.0, 14.0, 15.0, 16.0, 17.0, 18.0, 19.0, 20.0 \\
      7 & $[$N1$]$ \\
      8 & 6.0, 7.0, 8.0, 9.0, 10.0 \\
      \hline
      
      9 & $[$SOIL2$]$, $[$END$]$\\
      10 & $[$THICKNESS2$]$, $[$PAD$]$ \\
      11 & 1.0, 2.0, 3.0, 4.0, 5.0, 6.0, 7.0, 8.0, 9.0, 10.0, \\
         & 11.0, 12.0, 13.0, 14.0, 15.0, 16.0, 17.0, 18.0, 19.0, 20.0, $[$PAD$]$\\
      12 & $[$N2$]$, $[$PAD$]$ \\
      13 & 6.0, 7.0, 8.0, 9.0, 10.0, $[$PAD$]$ \\
      \hline
      
      14 & $[$SOIL3$]$, $[$END$]$, $[$PAD$]$ \\
      15 & $[$THICKNESS3$]$, $[$PAD$]$ \\
      16 & 1.0, 2.0, 3.0, 4.0, 5.0, 6.0, 7.0, 8.0, 9.0, 10.0, \\
         & 11.0, 12.0, 13.0, 14.0, 15.0, 16.0, 17.0, 18.0, 19.0, 20.0, $[$PAD$]$ \\
      17 & $[$N3$]$, $[$PAD$]$ \\
      18 &6.0, 7.0, 8.0, 9.0, 10.0, $[$PAD$]$ \\
      \hline
      
      19 & $[$SOIL4$]$, $[$END$]$, $[$PAD$]$ \\
      20 & $[$THICKNESS4$]$, $[$PAD$]$ \\
      21 & 1.0, 2.0, 3.0, 4.0, 5.0, 6.0, 7.0, 8.0, 9.0, 10.0, \\
         & 11.0, 12.0, 13.0, 14.0, 15.0, 16.0, 17.0, 18.0, 19.0, 20.0, $[$PAD$]$ \\
      22 & $[$N4$]$, $[$PAD$]$ \\
      23 & 6.0, 7.0, 8.0, 9.0, 10.0, $[$PAD$]$ \\
      \hline
      
      24 & $[$END$]$, $[$PAD$]$ \\
      \hline
    \end{tabular}
    \label{tab:TableS2}
\end{table}

\clearpage
\newgeometry{left=2.5cm,right=2.5cm,bottom=0.1cm,top=0.1cm} 
\restoregeometry

\clearpage
\subsubsection*{Training data generation}
\label{sec:TrainingDataGeneration}
Synthetic data was generated using a rock-physics model~\cite{Solazzi_etal_2021} modeling the elastic properties of soils under varying saturation conditions.
It extends the Hertz-Mindlin~\cite{Mindlin_1949} model using Bishop's~\cite{Bishop_Blight_1963} effective stress definition, to account for stiffness changes associated with capillary stresses.

\begin{figure}[p]
    \centering
    \includegraphics[width=\textwidth]{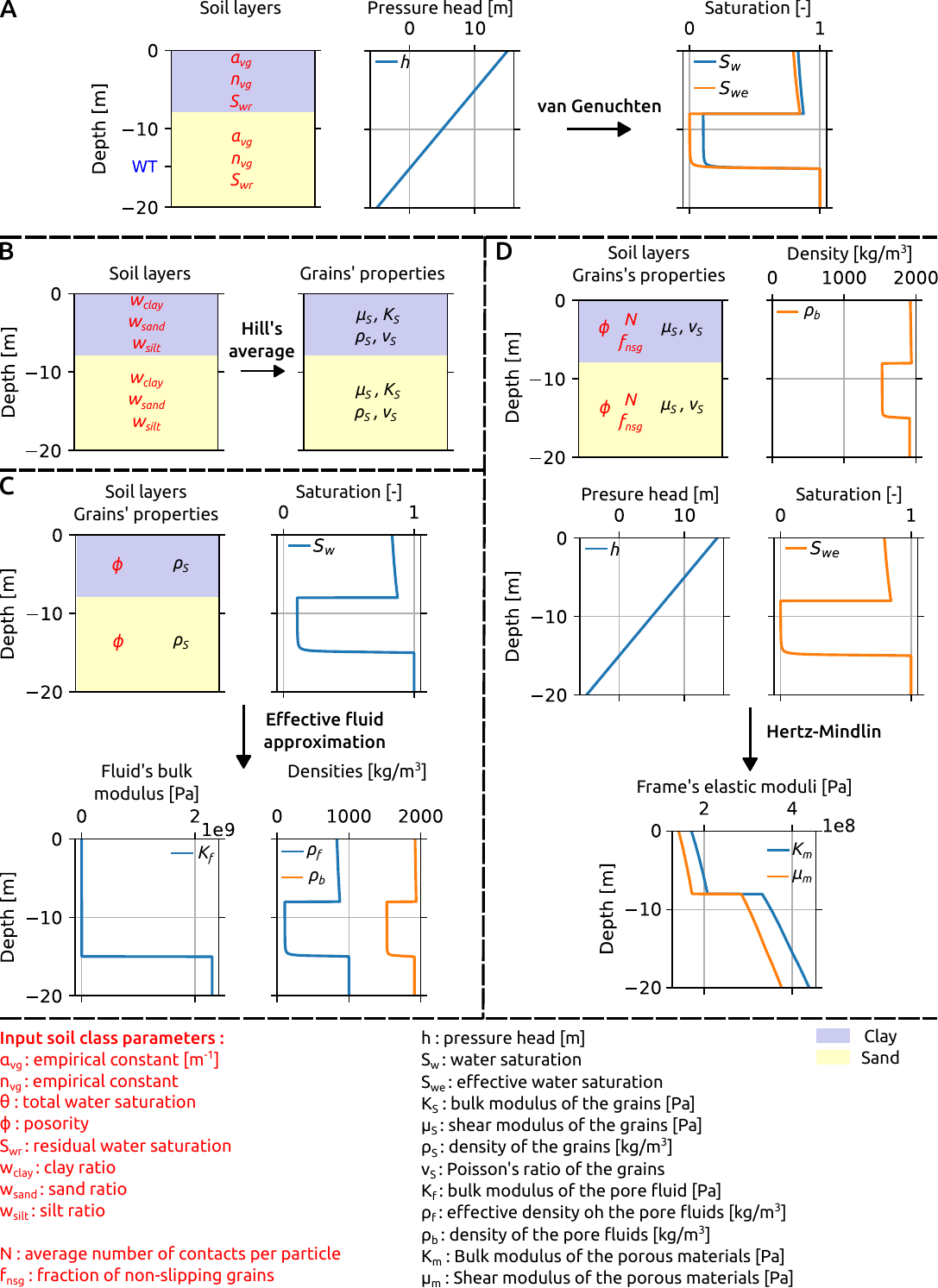}
    \caption{\textbf{Rock-physics model workflow for the computing of the frame's elastic moduli.}}
    \label{fig:FigureS2}
\end{figure}

However, the original model was developed to simulate DCs for a homogeneous soil with a given WT level, limiting its applicability to multi-layer geological contexts.
To address this limitation, we adapted the model for use with multiple soil layers, enabling more accurate modeling of complex subsurface structures (fig.~\ref{fig:FigureS2}).

\subsubsection*{Van Genuchten}
\label{sec:VanGenuchten}
We conceptualize a medium as a stack of $L$ layers, each composed of one soil type (clay, silt, loam, or sand), an average number of contacts per particle $N$, and a thickness $h$.
Each soil type is defined in table~\ref{tab:TableS3}, with constituent ratios of pure sand ($\gamma_{sand}$), clay ($\gamma_{clay}$), and silt ($\gamma_{silt}$), porosity ($\phi$), van Genuchten parameters ($\alpha_{vg}$, $n_{vg}$, and $\theta_{vg}$)~\cite{vanGenuchten_1980}, residual water saturation ($S_{wr}$), and the fraction of non-slipping grains ($f_{nsg}$).
As in the original paper, this medium of $L$ layers is discretized in a stack of $n$ cells of thickness $dz$ over a depth axis $z$.
Since our medium is not homogeneous, we can generalize its description with the parameters $\gamma_{sand}(z)$, $\gamma_{clay}(z)$, $\gamma_{silt}(z)$, $\phi(z)$, $\alpha_{vg}(z)$, $n_{vg}(z)$, $\theta_{vg}(z)$, $S_{wr}(z)$, and $N(z)$, as non continuous functions that vary with depth depending on the soil class.
The parameter $f_{nsg}$ remains constant at 0.3 in our application.

\begin{table}[!htbp]
\centering
  \caption{\textbf{Soil type petrophysical parameters.}
  Parameters describing each soil class~\cite{Solazzi_etal_2021} with with constituent ratios of pure sand ($\gamma_{sand}$), clay ($\gamma_{clay}$), and silt ($\gamma_{silt}$), porosity ($\phi$), van Genuchten parameters ($\alpha_{vg}$, $n_{vg}$, and $\theta_{vg}$), residual water saturation ($S_{wr}$), and the fraction of non-slipping grains ($f_{nsg}$).
  }
  \begin{tabular}{lccccccccc}
      \hline
      Soil type & $\gamma_{sand}$ & $\gamma_{clay}$ & $\gamma_{silt}$ & $\phi$ & $\alpha_{vg}$ $[cm^{-1}]$ & $n_{vg}$ & $\theta_{vg}$ & $S_{wr}$ & $f_{nsg}$ \\
      \hline
      Sand & 0.927 & 0.029 & 0.044 & 0.43 & 14.5 & 2.68 & 0.045 & 0.10 & 0.3 \\
      Loam & 0.400 & 0.197 & 0.403 & 0.43 & 3.6 & 1.56 & 0.078 & 0.18 & 0.3 \\
      Silt & 0.058 & 0.095 & 0.847 & 0.46 & 1.6 & 1.37 & 0.034 & 0.07 & 0.3 \\
      Clay & 0.149 & 0.552 & 0.299 & 0.38 & 0.8 & 1.09 & 0.068 & 0.18 & 0.3 \\
      \hline
    \end{tabular}
    \label{tab:TableS3}
\end{table}

We solve Richards' equation~\cite{Richards_1931}
\begin{equation}
    \frac{\partial}{\partial z}\left(\kappa (h) \frac{\partial}{\partial z}(h + z)\right) - \phi \frac{\partial S_w(h)}{\partial t} = 0 ,
    \label{eq:richards}
\end{equation}
where $\kappa (h)$ is the hydraulic conductivity, $S_w(h)$ is the water saturation, and $h$ is is the pressure head
\begin{equation}
    h = \frac{-p_c}{\rho_wg},
    \label{eq:h}
\end{equation}
with $p_c=p_a-p_w$ the capillary pressure, $p_a$ the air pressure, $p_w$ the water pressure, $\rho_w$ the water density, and $g$ the gravitational acceleration on Earth. 
Equation~\ref{eq:richards} assumes that the non-wetting phase (air) is infinitely mobile and, thus, $p_a$ variations in the subsurface are negligible.
Consequently, the pressure head can be approximated by $h=p_w/\rho_wg$.

We consider that the vadose zone is in equilibrium, that is, that the water is immobile.
Thus, $S_w$ does not vary with time and Equation~\ref{eq:richards} is reduced to
\begin{equation}
    \frac{\partial}{\partial z}\left(\kappa (h) \frac{\partial}{\partial z}(h + z)\right) = 0 .
\end{equation}
Additionally Darcy's flow is null, meaning that $\kappa (h) \frac{\partial}{\partial z}(h + z) = 0$.
Thus, the result of Equation 15 is given by $h=c-z$ with $c$ being an integration constant.
If we take $z=0$ at the surface, $h=0$ at the water table level $z=wt$, we have:
\begin{equation}
    h = wt-z.
    \label{eq:h2}
\end{equation}
This part remains unchanged from the original paper because $h$ is not a function of the soil type.

We replace $h$ in the Van Genuchten's~\cite{vanGenuchten_1980} constitutive model to relate the effective water saturation $S_{we}$ and $h$, and obtain
\begin{equation}
    S_{we}(z) =
    \begin{cases}
      \left\{ 1 + [\alpha_{vg}(z)(wt-z)]^{n_{vg}(z)}\right\} ^{-m_{vg}(z)}, & \text{for}\ z<wt, \\
      1, & \text{for}\ z \geq wt,
    \end{cases}
\end{equation}
where $\alpha_{vg}(z)$ denotes the inverse of the entry pressure, a value related to the air pressure needed to displace water from the larger pore sizes, and $n_{vg}(z)$ and $m_{vg}(z)=1-1/n_{vg}(z)$ are parameters related to the pore size distribution~\cite{Carsel_Parrish_1988}.
The water saturation $S_w$ is then obtained by
\begin{equation}
    S_w(z) = S_{we}(z) (1-S_{wr}(z)) + S_{wr}(z).
\end{equation}

\subsubsection*{Hill's average}
\label{sec:HillsAverage}
The elastic properties of the grains is computed using the Hill's averaging formula~\cite{Hill_1952} for each soil layer
\begin{equation}
    K_s(z) = \frac{1}{2} \left[ \sum _i ^m \gamma_i(z) K_{s,i}(z) + \left( \sum _i ^m \frac{\gamma_i(z)}{K_{s,i}(z)} \right) ^{-1} \right],
\end{equation}
\begin{equation}
    \mu_s(z) = \frac{1}{2} \left[ \sum _i ^m \gamma_i(z) \mu_{s,i}(z) + \left( \sum _i ^m \frac{\gamma_i(z)}{\mu_{s,i}(z)} \right) ^{-1} \right],
\end{equation}
where $m$ is the number of constituents (pure clay, sand, or silt), $\gamma_i$ is the corresponding volumetric fraction, and $K_{s,i}$ and $\mu_{s,i}$ are the bulk and shear moduli of the i-th constituent, respectively (table~\ref{tab:TableS4}).
The effective density of the solid grains is given by
\begin{equation}
    \rho_s(z) = \sum_{i=1} ^m \gamma_i(z) \rho_{s,i}(z),
\end{equation}
and the Poisson's ratio corresponds to
\begin{equation}
    \nu_s(z) = \frac{3K_s(z) - 2\mu_s(z)}{2(3K_s(z) +\mu_s(z))}.
\end{equation}

\begin{table}[!htbp]
\centering
  \caption{\textbf{Soil constituents mechanical parameters.}
  Bulk moduli $K_{s}$, shear moduli $\mu_{s}$, and density $\rho_{s}$ of each soil constituent type (pure sand, clay, and silt).
  }
  \begin{tabular}{lccc}
      \hline
      Soil constituent & $K_{s}$ $[GPa]$ & $\mu_{s}$ $[GPa]$ & $\rho_{s}$ $[kg/m^3]$ \\
      \hline
      Pure sand & 45 & 25 & 2580 \\
      Pure silt & 45 & 37 & 2600 \\
      Pure clay & 6.8 & 25 & 2600 \\
      \hline
    \end{tabular}
    \label{tab:TableS4}
\end{table}

\subsubsection*{Biot-Gassmann-Wood}
\label{sec:BiotGassmanWood}
The Biot-Gassmann low-frequency relationships~\cite{Biot_1962,Gassmann_1951,Mavko_etal_2009} are used to account for the effects of a saturating fluid on the elastic moduli of porous materials
\begin{equation}
    K(z,S_w) = K_m(z,S_w) + \frac{\left( 1 - \frac{K_m(z,S_w)}{K_s(z)} \right)^2}{\frac{\phi(z)}{K_f(S_w)} - \frac{1-\phi(z)}{K_s(z)} - \frac{K_m(z,S_w)}{K_s(z)^2}} ,
    \label{eq:K}
\end{equation}
\begin{equation}
    \mu(z,S_w) = \mu_m(z,S_w),
    \label{eq:mu}
\end{equation}
where $K_m$ and $\mu_m$ are the drained moduli of the porous medium, $\phi$ is the porosity, and $K_f$ is the fluid bulk modulus approximated by
\begin{equation}
    K_f(S_w) = \left[ \frac{S_w(z)}{K_w} + \frac{1-S_w(z)}{K_a} \right]^{-1},
\end{equation}
with $K_w$ and $K_a$ denoting the bulk moduli of water and air, respectively.
Similarly, the effective fluid density is given by
\begin{equation}
    \rho_f(S_w) = S_w(z) \rho_w + \rho_a (1 - S_w(z)),
\end{equation}
and the effective bulk density by
\begin{equation}
    \rho_b(S_w) = (1-\phi(z))\rho_s(z) + \phi(z)[S_w(z) \rho_w + (1-S_w(z))\rho_a],
\end{equation}
where $\rho_a$ and $\rho_s$ denote the density of air and of the solid grains, respectively.

\subsubsection*{Hertz-Mindlin}
\label{sec:HertzMindlin}
We use a modified Hertz-Mindlin model~\cite{Solazzi_etal_2021} to to estimate the effective elastic properties of the soil's frame, the moduli of the porous medium $K_m$ and $\mu_m$
\begin{equation}
    K_m(z,S_w) = \left[ \frac{N^2(1-\phi(z))^2\mu_s(z)^2}{18\pi^2(1-\nu_s(z))^2} P_e(z,S_w) \right]^{1/3},
\end{equation}
\begin{equation}
    \mu_m(z,S_w) = \frac{2+3f_{nsg}-(1+3f_{nsg})\nu_s(z)}{5(2-\nu_s(z))} \left[ \frac{3N(z)^2(1-\phi(z))^2\mu_s(z)^2}{2\pi^2(1-\nu_s(z))^2} P_e(z,S_w) \right]^{1/3},
\end{equation}
where $N$ is the average number of contacts per particle, $f_{nsg}$ the fraction of non-slipping grains, and $P_e(z,S_w)$ is the saturation- and depth-dependent effective stress accounting for the capillary suction effect~\cite{Bishop_Blight_1963}
\begin{equation}
    P_e(S_w) = \sigma(S_w) - p_a + \sigma_s(S_w),
\end{equation}
where $\sigma = \rho_b(S_w)gz$ is the overburden stress and $\sigma_s$ is the capillary suction stress
\begin{equation}
    \sigma_s(S_w)=
    \begin{cases}
      S_{we} \rho_w  g  h, & \text{for}\ h>0, \\
      0, & \text{for}\ h \leq 0.
    \end{cases} 
\end{equation}

\clearpage
\newgeometry{left=2.5cm,right=2.5cm,bottom=0.1cm,top=0.1cm} 
\restoregeometry

\subsubsection*{Seismic wave velocities}
\label{sec:SeismicWaveVelocities}
\begin{figure}[p]
    \centering
    \includegraphics[width=\textwidth]{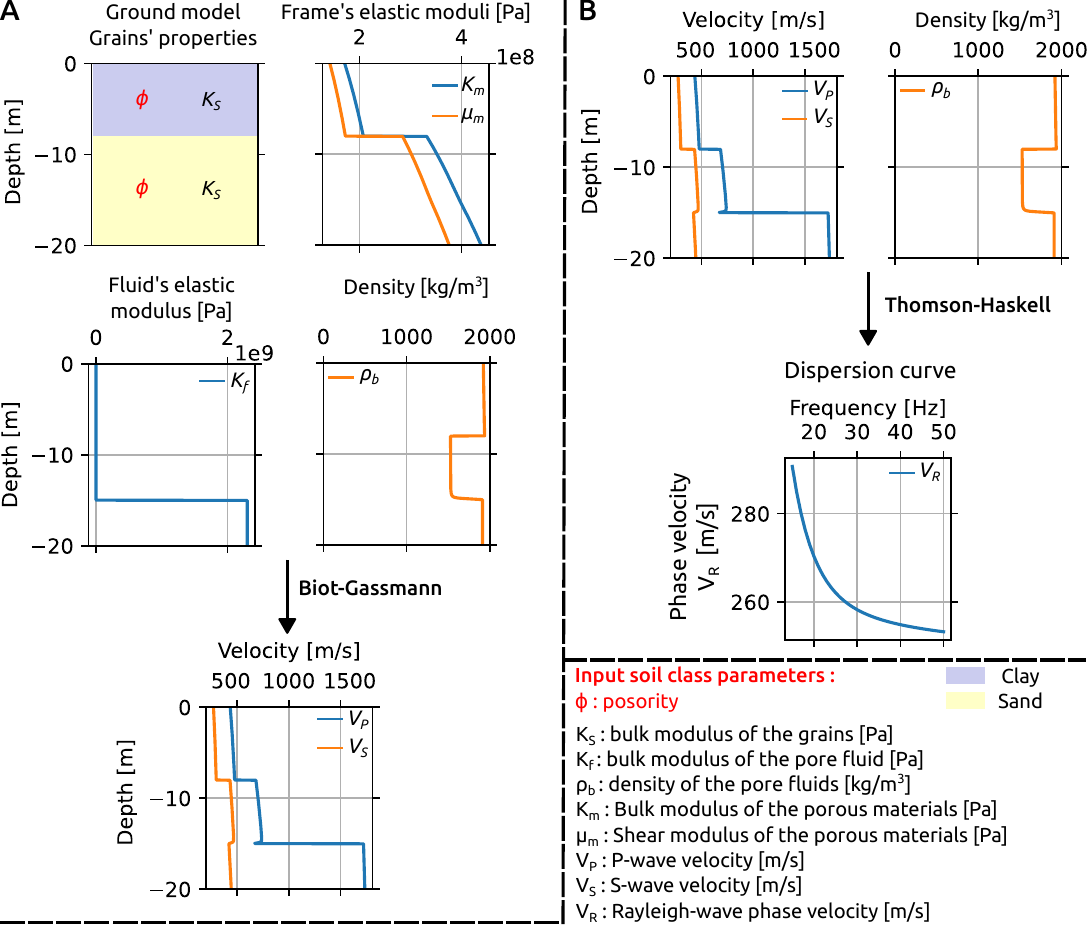}
    \caption{\textbf{Rock-physics model workflow for the computing of seismic wave velocities.}}
    \label{fig:FigureS3}
\end{figure}

Seismic wave velocities are computed from the elastic properties obtained through the rock-physics model (fig.~\ref{fig:FigureS3}).
We use the results of the Biot-Gassmann equations~\ref{eq:K} and \ref{eq:mu} to compute the P- and S- wave velocities over depth such as
\begin{equation}
    V_{P}(z) = \sqrt{\frac{K(z) + \frac{4}{3}\mu(z)}{\rho_{b}(z)}},
\end{equation}
\begin{equation}
    V_{S}(z) = \sqrt{\frac{\mu(z)}{\rho_{b}(z)}}.
\end{equation}

Finally, we use the classical Thomson-Haskell matrix propagator technique~\cite{Thomson_1950,Haskel_1953} in the stack of $n$ layers, plus two additional custom sub-layers to better approximate real cases with a rigid substratum (table~\ref{tab:TableS5}), to compute the DC of the Rayleigh-wave.
Note that we only use the fundamental mode.

\begin{table}[!htbp]
\centering
  \caption{\textbf{Sub-layers added to the rock-physics model.}
  These layers are added beneath the velocity ground model obtained with the rock-physics model.
  The last layer corresponds to a half-space.
  }
    \begin{tabular}{ccccc}
      \hline
      Sub-layer & Thickness $[m]$ & $V_P$ $[m/s]$ & $V_S$ $[m/s]$ & $\rho$ $[kg/m^3]$ \\
      \hline
      1 & 10 & 1500 & 750 & 2000 \\
      half-space & $\infty$ & 4000 & 2000 & 2500 \\
      \hline
    \end{tabular}
    \label{tab:TableS5}
\end{table}

\clearpage

\clearpage
\subsubsection*{Modeling parameters}
\label{sec:ModelingParameters}
Each generated synthetic model was parameterized with a random number of one to four soil layers.
Each layer had a random thickness ranging from 1 to 20~m with a step of 1~m, constrained such that the total thickness of all layers was exactly 20~m, a random soil type and a random $N$ value ranging from 6 to 10, in increments of 1.
The WT level was also randomly assigned a value between 0.5 and 10~m, in increments of 0.5~m.
Note that the parameter space comprises a total of 1,323,654,400 different combinations.

\subsubsection*{Model training}
\label{sec:ModelTraining}
\textsc{Silex} was trained in two phases.
Initially, it was trained on a dataset of 1,166,446 synthetic sample pairs, consisting of soil descriptions and their corresponding modeled DC values, normalized between 0 and 1.
This dataset covered only 0.9\% of the total 1,323,654,400 possible soil, N, thicknesses, and WT level combinations.
The model was trained for 100 epochs with a batch size of 64 samples, using a validation set of 100,000 samples for early stopping and a test set of 50,000 samples to evaluate final performance.
We used the \textit{Root Mean Square Propagation} optimizer (learning rate of $10^{-3}$), with the \textit{sparse categorical cross-entropy} as the loss function.
\textsc{Silex} achieved an accuracy of 92.2\% on the training data, 91.7\% on the validation set, and 91\% on the test set.

To enhance robustness on real data, \textsc{Silex} was retrained for an additional 5 epochs using the same dataset, but with Gaussian noise added to the input DCs to introduce variability.
Although accuracy dropped to 83\% on the training set and 82\% on the test set, this approach aims to improve the model's generalization.

\subsubsection*{Model evaluation}
\label{sec:ModelEvaluation}
DCs were recomputed from \textsc{Silex}'s inferences using the same rock-physics model~\cite{Solazzi_etal_2021}, allowing for a direct comparison with the input DCs.
On average between September 4, 2020, and September 3, 2023, the root mean square error (RMSE) on the DCs was initially estimated at 38 m/s (or a normalized RMSE of 26\%), and at 12 m/s (8\%) after retraining.
The RMSE is defined as
\begin{equation}
    RMSE(V_R, \hat{V_R}) = \frac{1}{n_{freqs}} \sum_{i=1}^{n_{freqs}}{\left( {V_R}_i - \hat{{V_R}_i} \right )^2}
    \ ,
    \label{eq:RMSE}
\end{equation}
where $V_R$ and $\hat{V_R}$ are the observed and predicted values, respectively.
The normalized RMSE (or NRMSE) is defined as
\begin{equation}
    NRMSE(V_R, \hat{V_R}) = \frac{ RMSE(V_R, \hat{V_R})}{{V_R}_{min} - {V_R}_{max}}
    \ .
    \label{eq:NRMSE}
\end{equation}
where ${V_R}_{min}$ and ${V_R}_{max}$ the minimum and maximum Rayleigh-wave phase velocity, respectively, of the DC between 15 and 50~Hz.

\begin{figure}[p]
    \centering
    \includegraphics[width=\textwidth]{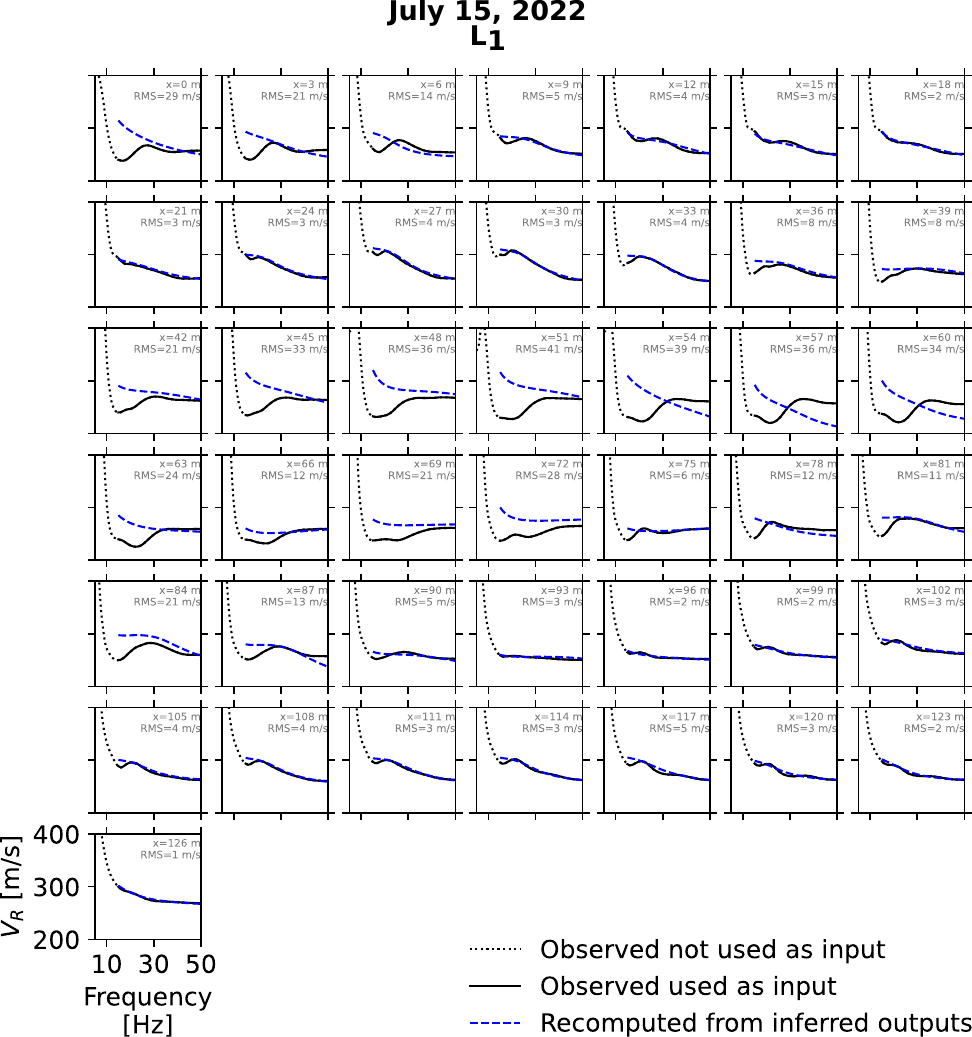}
    \caption{\textbf{Comparison between the real and the inferred dispersion curves along geophone line $L_1$ on July 15, 2022.} 
    Shown are the input dispersion curves (black) compared with those recomputed using the rock-physics model based on \textsc{Silex} outputs (blue).
    Only frequencies between 15 and 50~Hz of the input dispersion curves are used by \textsc{Silex}, so the recomputed curves are constrained to this range.
    Positions along the x-axis and root mean square errors (RMSE) are indicated in grey.
    }
    \label{fig:FigureS4}
\end{figure}

\begin{figure}[p]
    \centering
    \includegraphics[width=\textwidth]{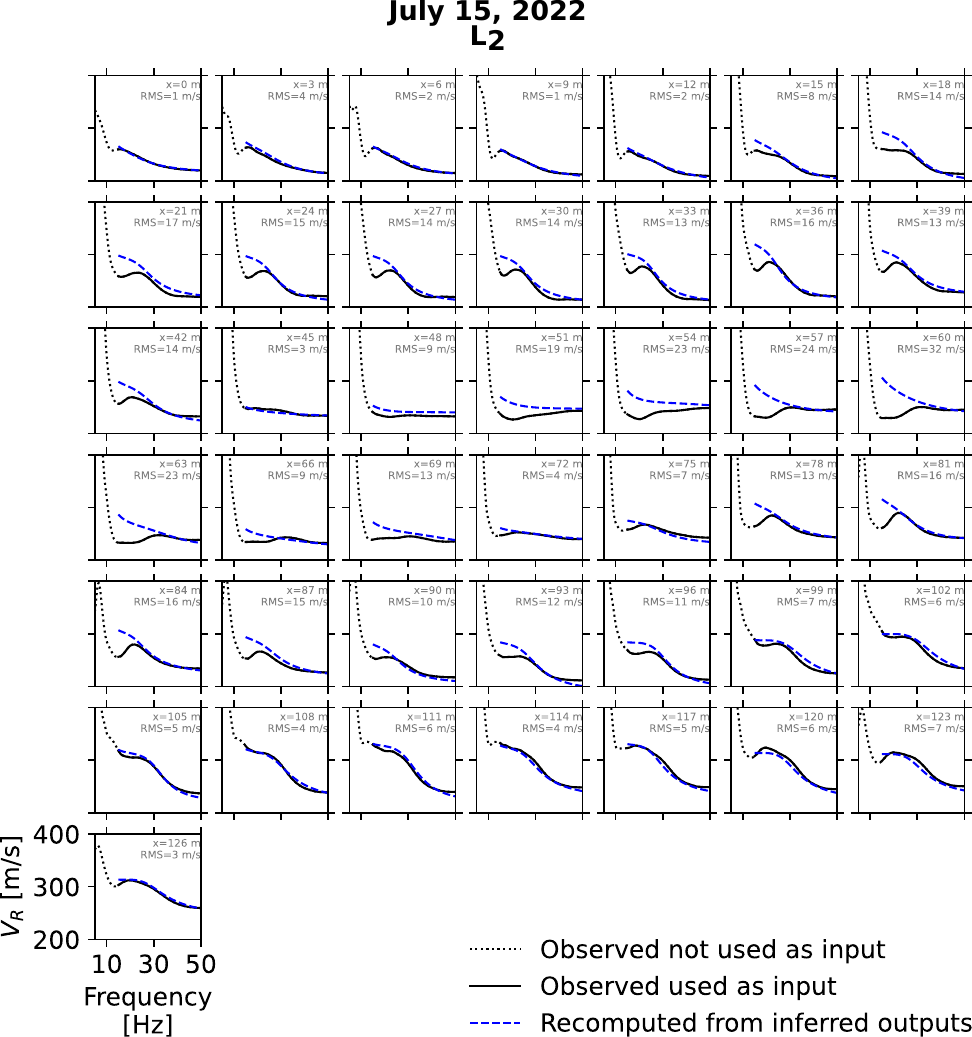}
    \caption{\textbf{Comparison between the real and the inferred dispersion curves along geophone line $L_2$ on July 15, 2022.} 
    Shown are the input dispersion curves (black) compared with those recomputed using the rock-physics model based on \textsc{Silex} outputs (blue).
    Only frequencies between 15 and 50~Hz of the input dispersion curves are used by \textsc{Silex}, so the recomputed curves are constrained to this range.
    Positions along the x-axis and root mean square errors (RMSE) are indicated in grey.
    }
    \label{fig:FigureS5}
\end{figure}

\begin{figure}[p]
    \centering
    \includegraphics[width=\textwidth]{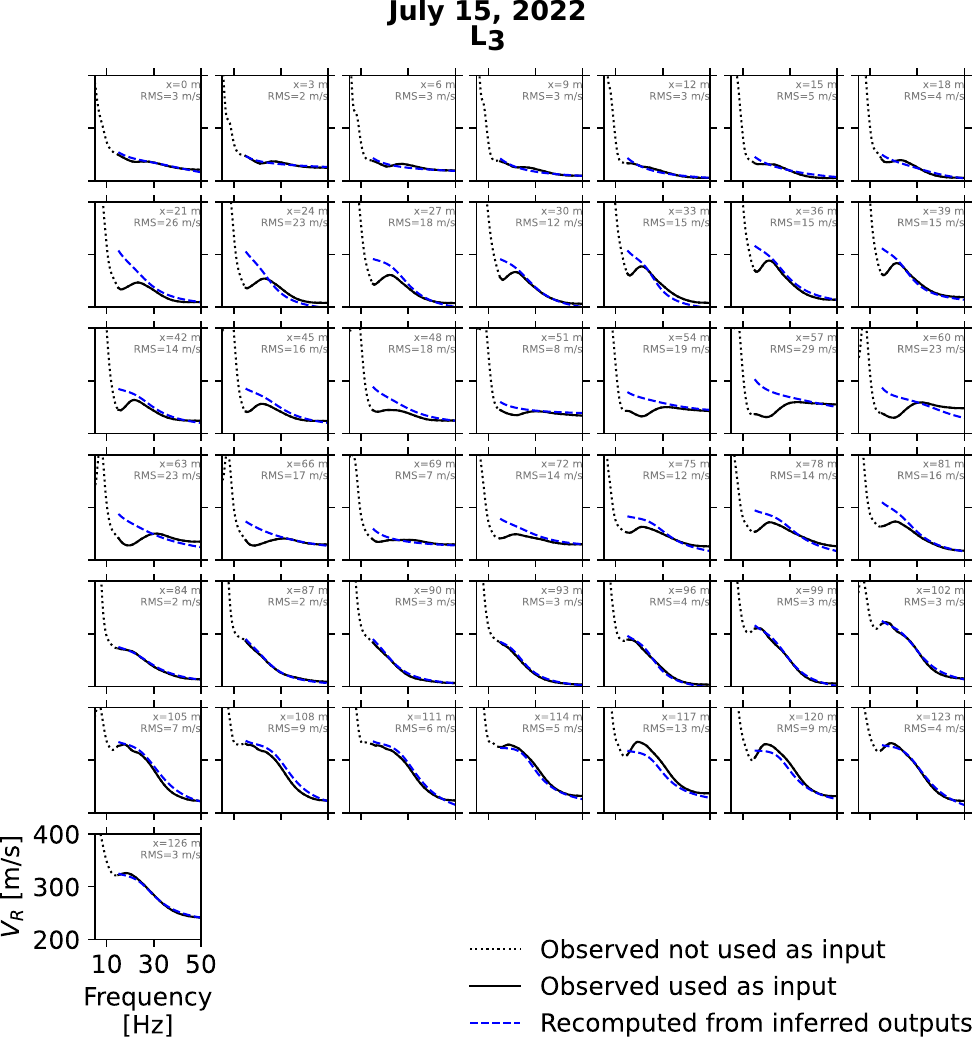}
    \caption{\textbf{Comparison between the real and the inferred dispersion curves along geophone line $L_3$ on July 15, 2022.} 
    Shown are the input dispersion curves (black) compared with those recomputed using the rock-physics model based on \textsc{Silex} outputs (blue).
    Only frequencies between 15 and 50~Hz of the input dispersion curves are used by \textsc{Silex}, so the recomputed curves are constrained to this range.
    Positions along the x-axis and root mean square errors (RMSE) are indicated in grey.
    }
    \label{fig:FigureS6}
\end{figure}

\begin{figure}[p]
    \centering
    \includegraphics[width=\textwidth]{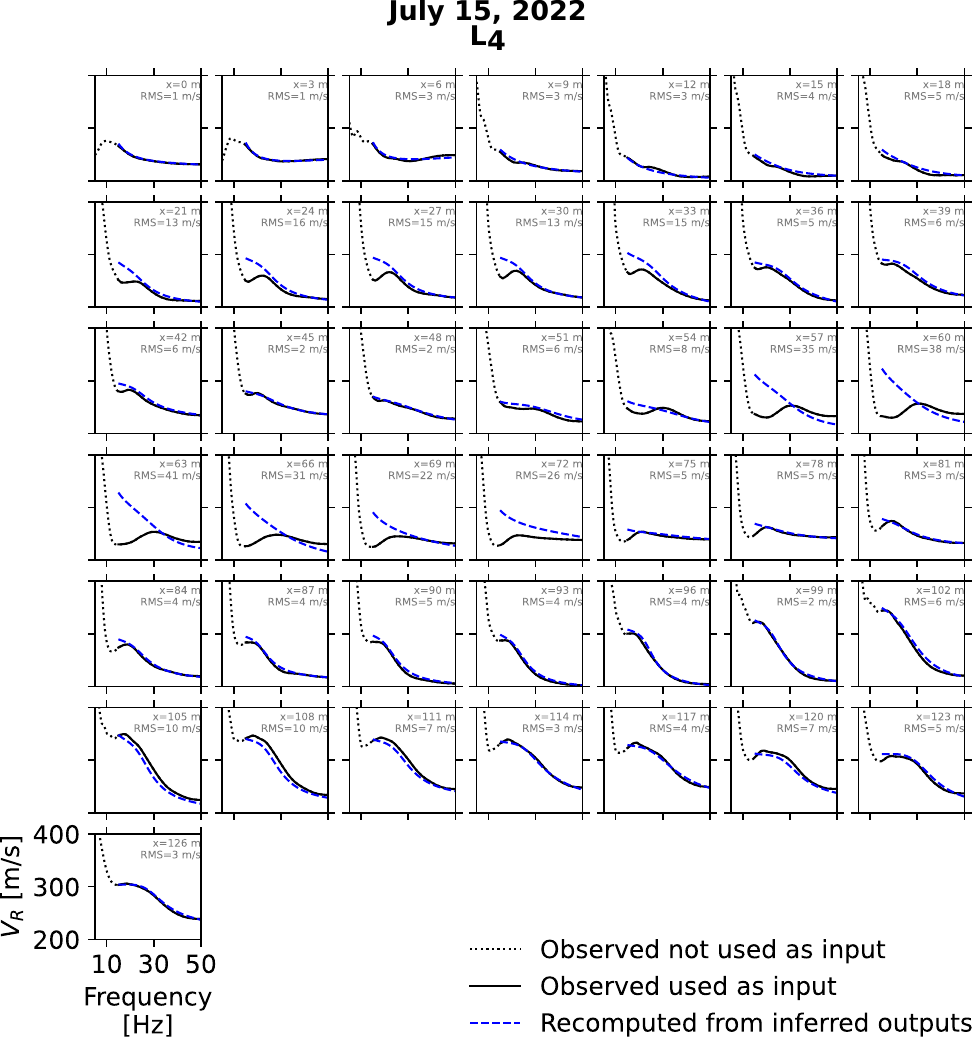}
    \caption{\textbf{Comparison between the real and the inferred dispersion curves along geophone line $L_4$ on July 15, 2022.} 
    Shown are the input dispersion curves (black) compared with those recomputed using the rock-physics model based on \textsc{Silex} outputs (blue).
    Only frequencies between 15 and 50~Hz of the input dispersion curves are used by \textsc{Silex}, so the recomputed curves are constrained to this range.
    Positions along the x-axis and root mean square errors (RMSE) are indicated in grey.
    }
    \label{fig:FigureS7}
\end{figure}

\begin{figure}[p]
    \centering
    \includegraphics[width=\textwidth]{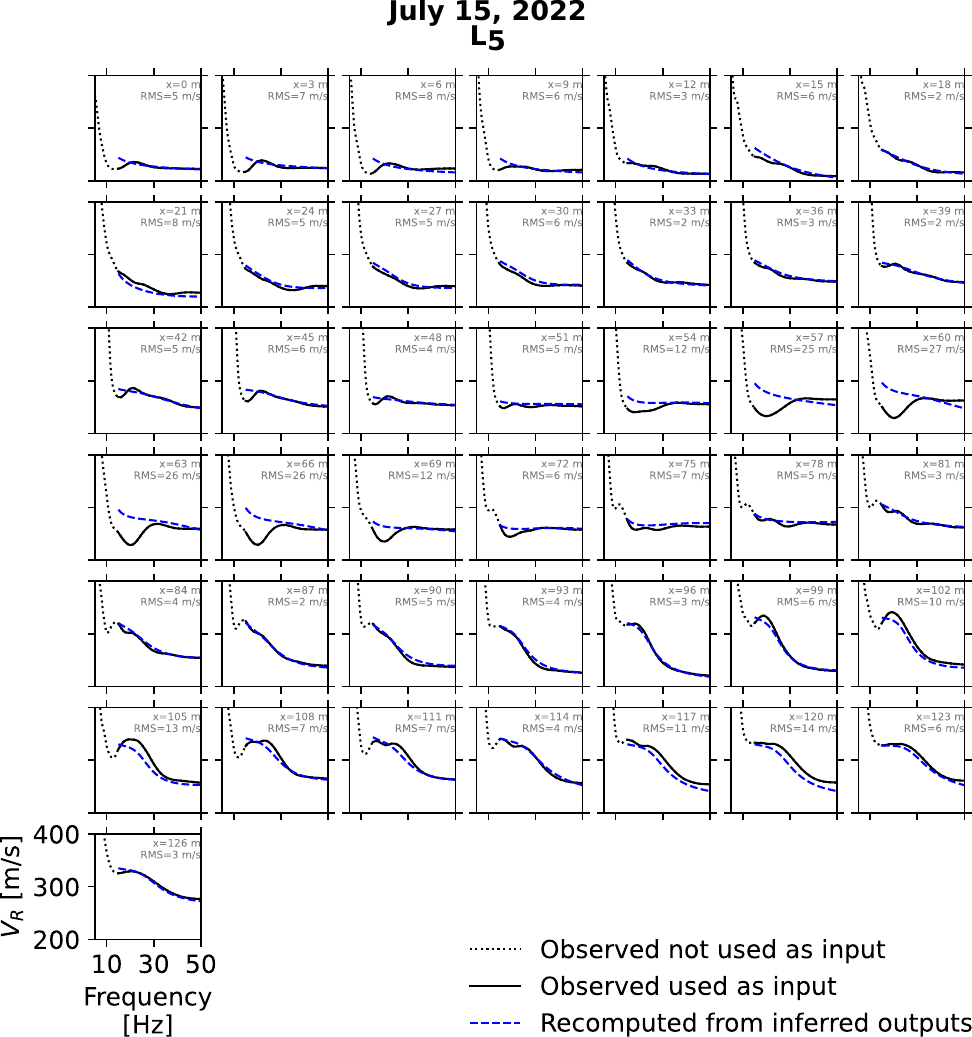}
    \caption{\textbf{Comparison between the real and the inferred dispersion curves along geophone line $L_5$ on July 15, 2022.} 
    Shown are the input dispersion curves (black) compared with those recomputed using the rock-physics model based on \textsc{Silex} outputs (blue).
    Only frequencies between 15 and 50~Hz of the input dispersion curves are used by \textsc{Silex}, so the recomputed curves are constrained to this range.
    Positions along the x-axis and root mean square errors (RMSE) are indicated in grey.
    }
    \label{fig:FigureS8}
\end{figure}

\begin{figure}[p]
    \centering
    \includegraphics[width=\textwidth]{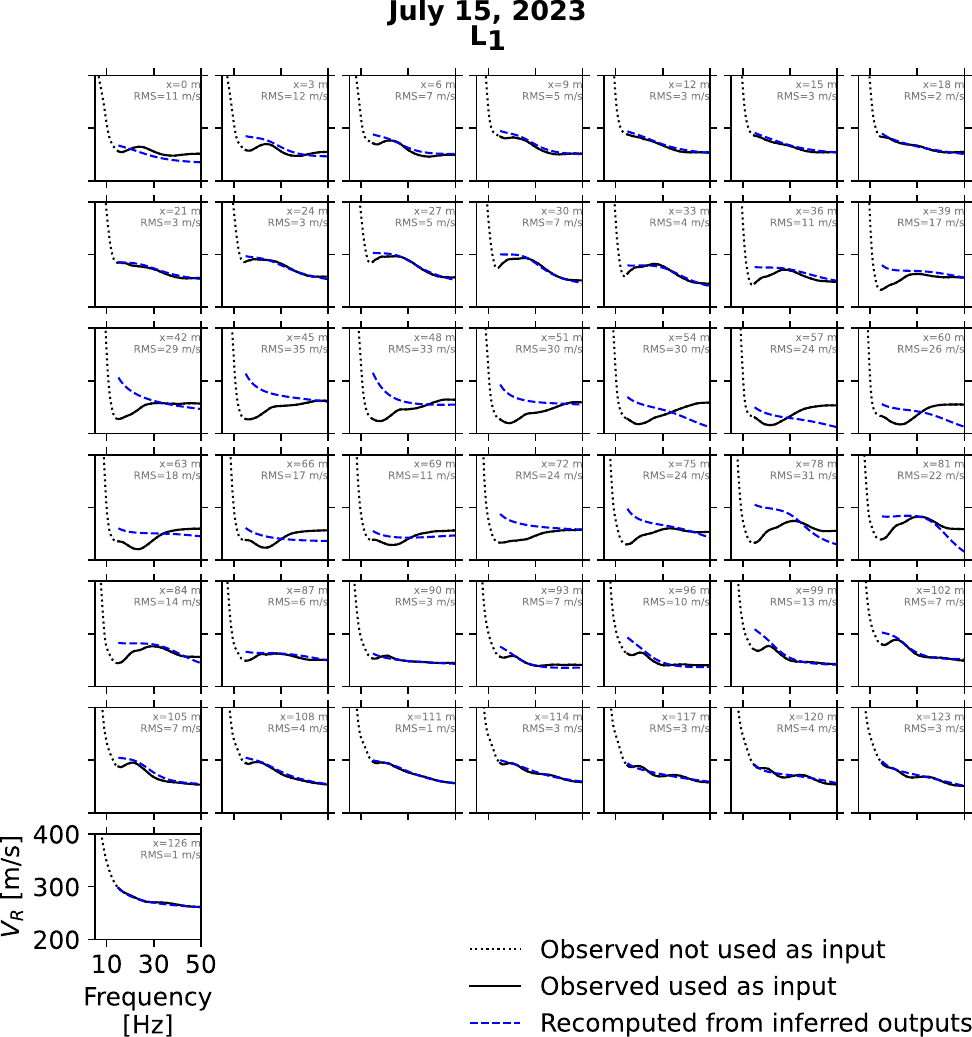}
    \caption{\textbf{Comparison between the real and the inferred dispersion curves along geophone line $L_1$ on July 15, 2023.} 
    Shown are the input dispersion curves (black) compared with those recomputed using the rock-physics model based on \textsc{Silex} outputs (blue).
    Only frequencies between 15 and 50~Hz of the input dispersion curves are used by \textsc{Silex}, so the recomputed curves are constrained to this range.
    Positions along the x-axis and root mean square errors (RMSE) are indicated in grey..
    }
    \label{fig:FigureS9}
\end{figure}

\begin{figure}[p]
    \centering
    \includegraphics[width=\textwidth]{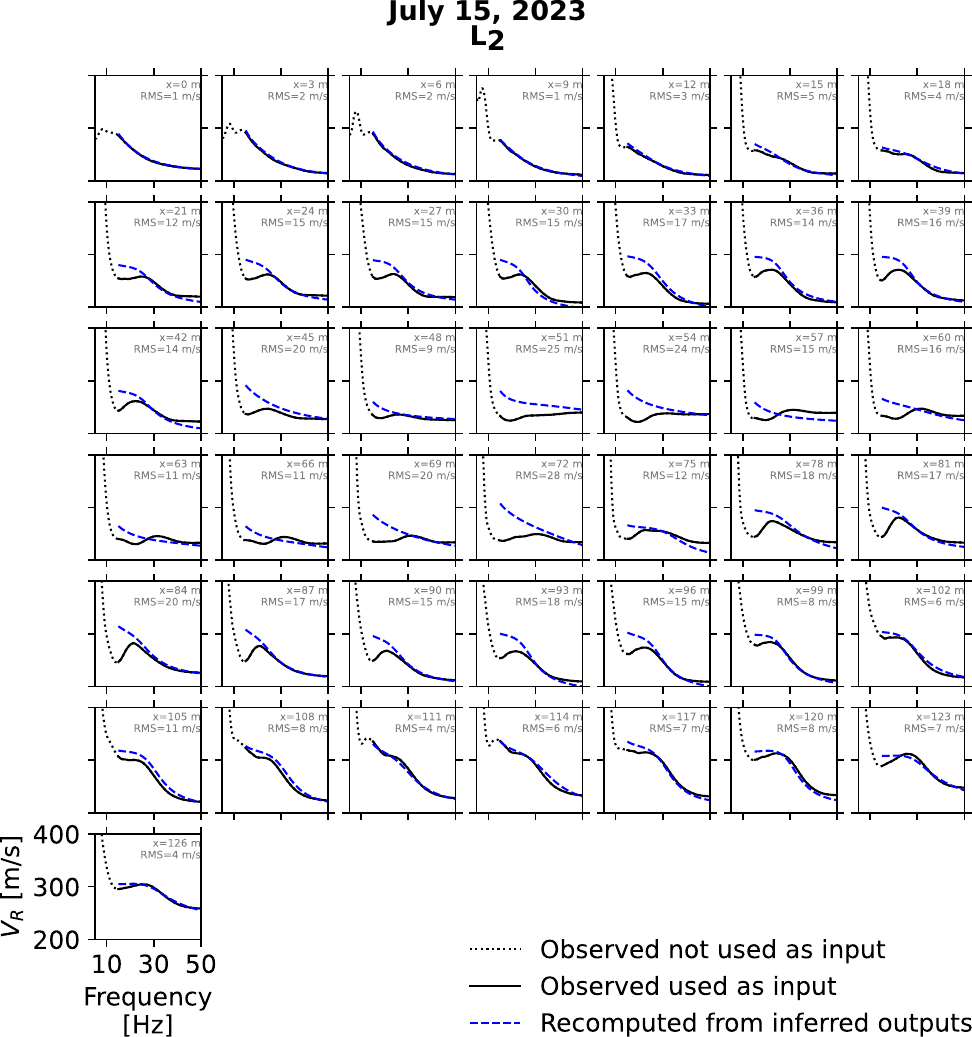}
    \caption{\textbf{Comparison between the real and the inferred dispersion curves along geophone line $L_2$ on July 15, 2023.} 
    Shown are the input dispersion curves (black) compared with those recomputed using the rock-physics model based on \textsc{Silex} outputs (blue).
    Only frequencies between 15 and 50~Hz of the input dispersion curves are used by \textsc{Silex}, so the recomputed curves are constrained to this range.
    Positions along the x-axis and root mean square errors (RMSE) are indicated in grey.
    }
    \label{fig:FigureS10}
\end{figure}

\begin{figure}[p]
    \centering
    \includegraphics[width=\textwidth]{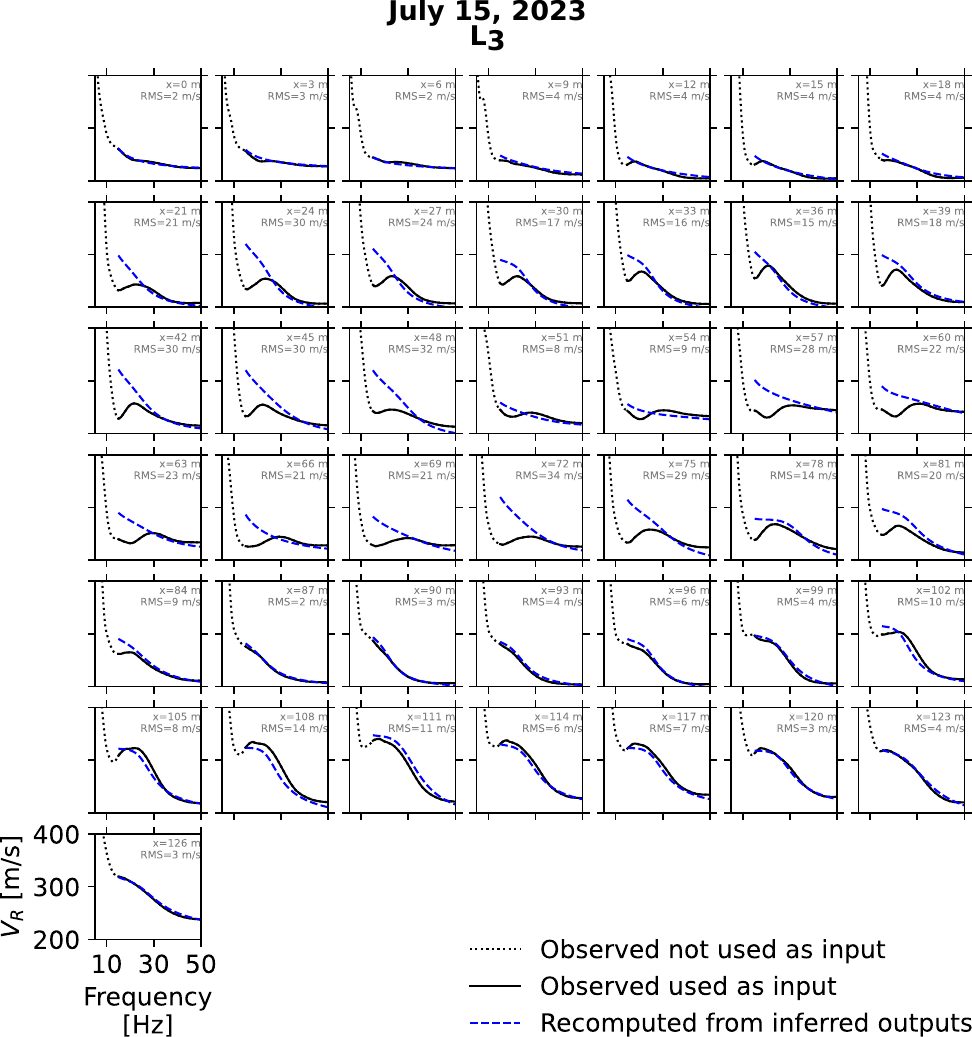}
    \caption{\textbf{Comparison between the real and the inferred dispersion curves along geophone line $L_3$ on July 15, 2023.} 
    Shown are the input dispersion curves (black) compared with those recomputed using the rock-physics model based on \textsc{Silex} outputs (blue).
    Only frequencies between 15 and 50~Hz of the input dispersion curves are used by \textsc{Silex}, so the recomputed curves are constrained to this range.
    Positions along the x-axis and root mean square errors (RMSE) are indicated in grey. 
    }
    \label{fig:FigureS11}
\end{figure}

\begin{figure}[p]
    \centering
    \includegraphics[width=\textwidth]{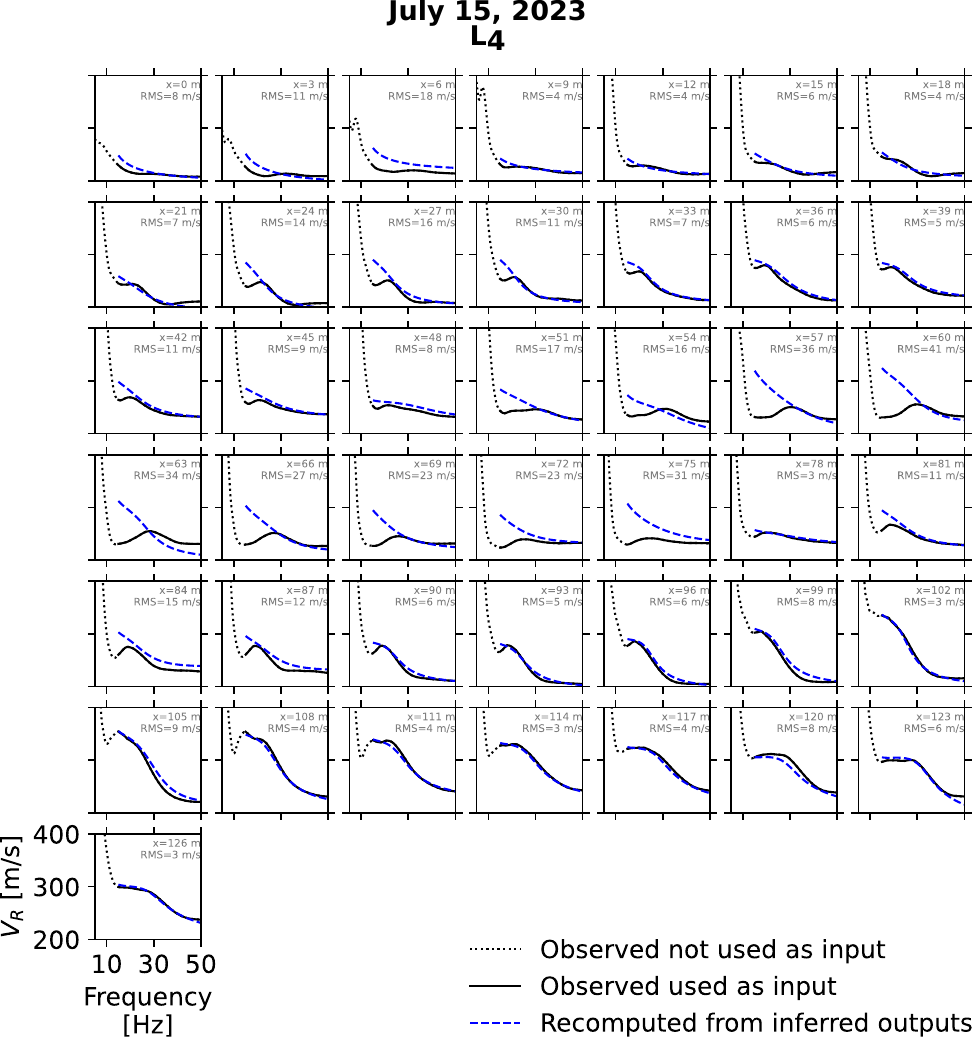}
    \caption{\textbf{Comparison between the real and the inferred dispersion curves along geophone line $L_4$ on July 15, 2023.} 
    Shown are the input dispersion curves (black) compared with those recomputed using the rock-physics model based on \textsc{Silex} outputs (blue).
    Only frequencies between 15 and 50~Hz of the input dispersion curves are used by \textsc{Silex}, so the recomputed curves are constrained to this range.
    Positions along the x-axis and root mean square errors (RMSE) are indicated in grey.
    }
    \label{fig:FigureS12}
\end{figure}

\begin{figure}[p]
    \centering
    \includegraphics[width=\textwidth]{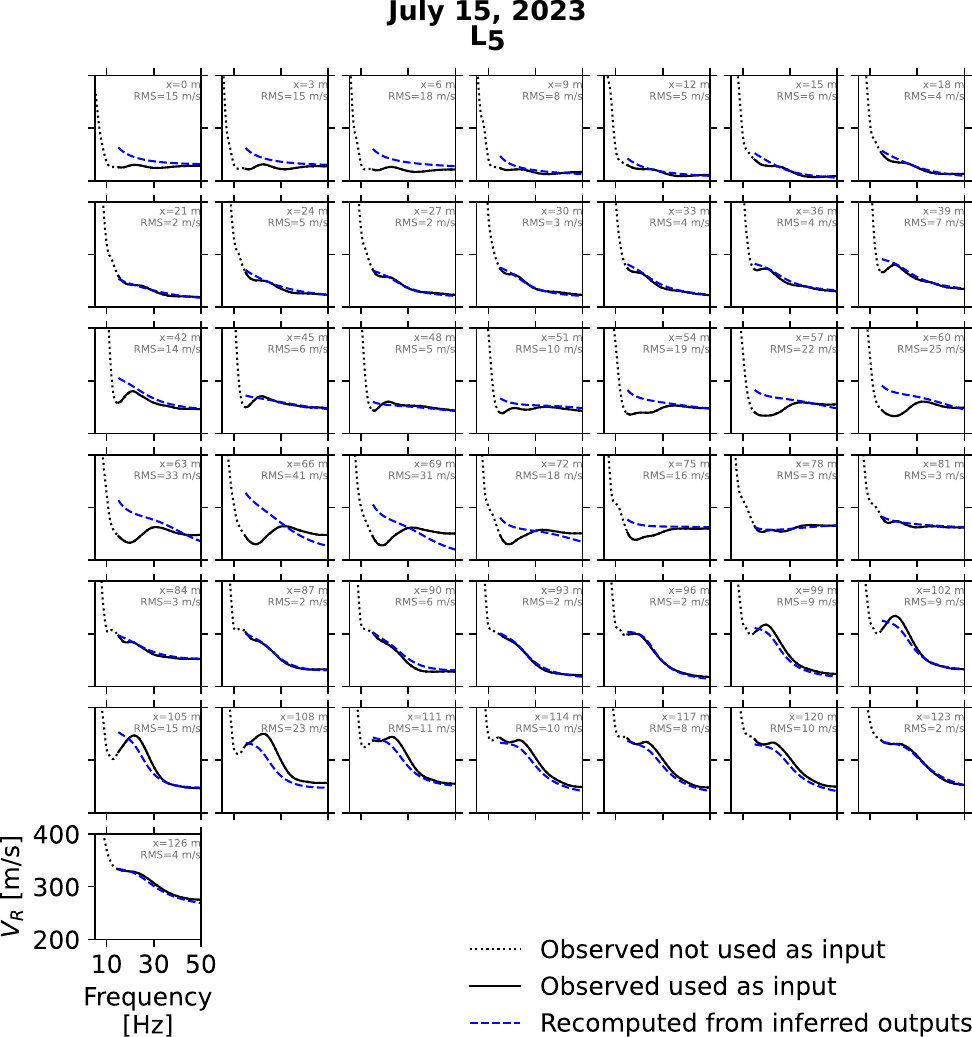}
    \caption{\textbf{Comparison between the real and the inferred dispersion curves along geophone line $L_5$ on July 15, 2023.} 
    Shown are the input dispersion curves (black) compared with those recomputed using the rock-physics model based on \textsc{Silex} outputs (blue).
    Only frequencies between 15 and 50~Hz of the input dispersion curves are used by \textsc{Silex}, so the recomputed curves are constrained to this range.
    Positions along the x-axis and root mean square errors (RMSE) are indicated in grey.
    }
    \label{fig:FigureS13}
\end{figure}

As examples, figs~\ref{fig:FigureS4} to \ref{fig:FigureS13} show the comparison between the input and recomputed DCs on July~15, 2022 and 2023.
The DCs marked in black were used as input data, while the red curves were recomputed using on the rock-physics model on \textsc{Silex}'s outputs.
Notably, the frequencies used as input were restricted to the range of 15 to 50~Hz, which is also reflected in the recomputed curves.
The RMSE, indicated in grey, quantify the misfit between the input and recomputed curves. 
Lower RMSE values are globally observed, demonstrating the model's efficacy, while some discrepancies appear at some positions (e.g. at $x$ = 54~m on fig.~\ref{fig:FigureS4}).
These inconsistencies highlight the model's limitations in approximating DCs with abrupt variations, which are inherent to the rock-physics model used to generate the training data.
Additionally, at these specific locations, the input DCs tend to be of lower quality and may suffer from mode contamination.
This occurs when a higher propagating mode is mistakenly identified as the fundamental mode (notably after 30~Hz in this example) during the automatic dispersion image picking process~\cite{Tarnus_etal_2022a,Tarnus_etal_2022b}, resulting in a bump in the DC.
Such contamination further complicates \textsc{Silex}'s ability to accurately capture the true subsurface characteristics at these points.
It is interesting to note that \textsc{Silex} seems to produce an average DC that attempts to balance the fit between the 15–30~Hz and 30–50~Hz frequency ranges.

\clearpage

\clearpage

\clearpage

\clearpage

\clearpage

\clearpage

\clearpage

\clearpage

\clearpage

\clearpage

\clearpage
\subsection*{Supplementary text}

\subsubsection*{Results' sectional views}
\begin{figure}[htbp]
    \centering
    \includegraphics[width=\textwidth]{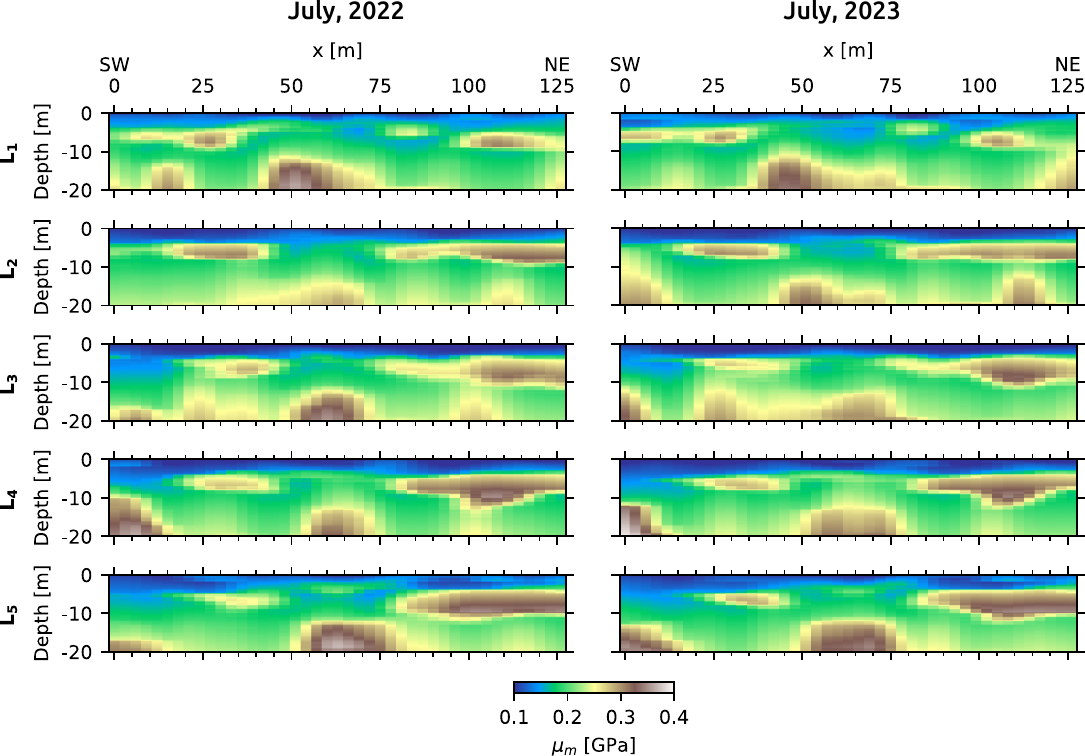}
    \caption{\textbf{Shear modulus sections.}
    Shear modulus ($\mu_m$) sections across the five geophone lines ($L_1$ to $L_5$) computed from the petrophysical inversion results using Hertz-Mindlin model~\cite{Mindlin_1949}, on July 2022 and 2023.
    }
    \label{fig:FigureS14}
\end{figure}

\begin{figure}[p]
    \centering
    \includegraphics[width=\textwidth]{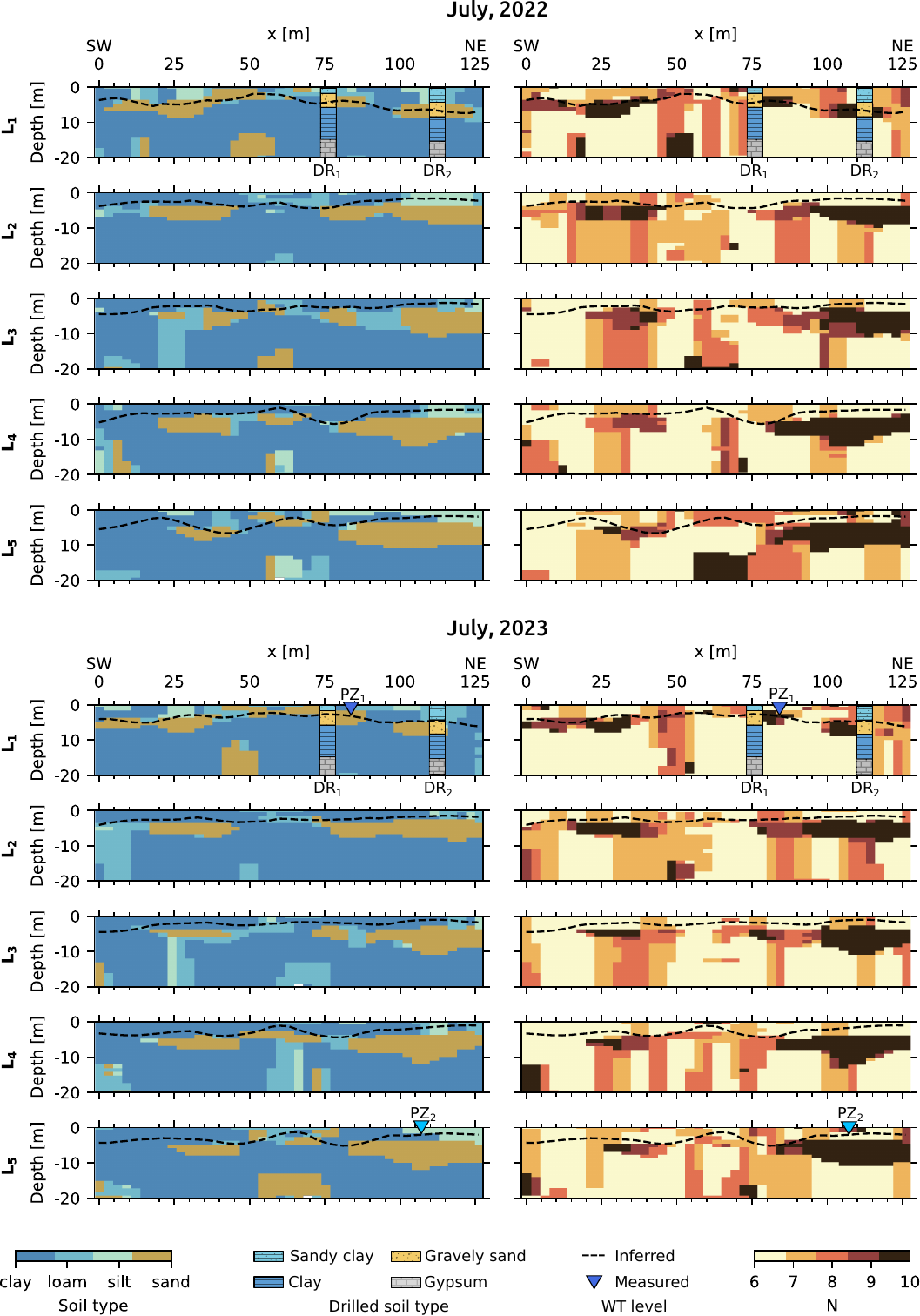}
    \caption{\textbf{Petrophysical inversion result sections.}
    Section representation of the petrophysical inversion results generated by the language model, across the five geophone lines ($L_1$ to $L_5$), on July 2022 and 2023.
    \textbf{Left:} Sections of soil types and water table (WT) level.
    \textbf{Right:} Sections of the average number of contacts per particles ($N$) and WT level.
    }
    \label{fig:FigureS15}
\end{figure}

Offering additional insights, figs~\ref{fig:FigureS14} and \ref{fig:FigureS15} break down the 3D~representations shown in the main text, in Figs~4 and 3, respectively, into individual 2D~sections.
The monthly average is calculated by counting the daily occurrences of each soil type and $N$ at every $(x,z)$ position, and then selecting the ones with the maximum occurrence
The average WT levels and $\mu_m$ are calculated as simple arithmetic means over a month.

\clearpage
\newgeometry{left=2.5cm,right=2.5cm,bottom=0.1cm,top=0.1cm} 
\restoregeometry

\subsubsection*{Variability}
\label{sec:variability}
\begin{figure}[hp]
    \centering
    \includegraphics[width=\textwidth]{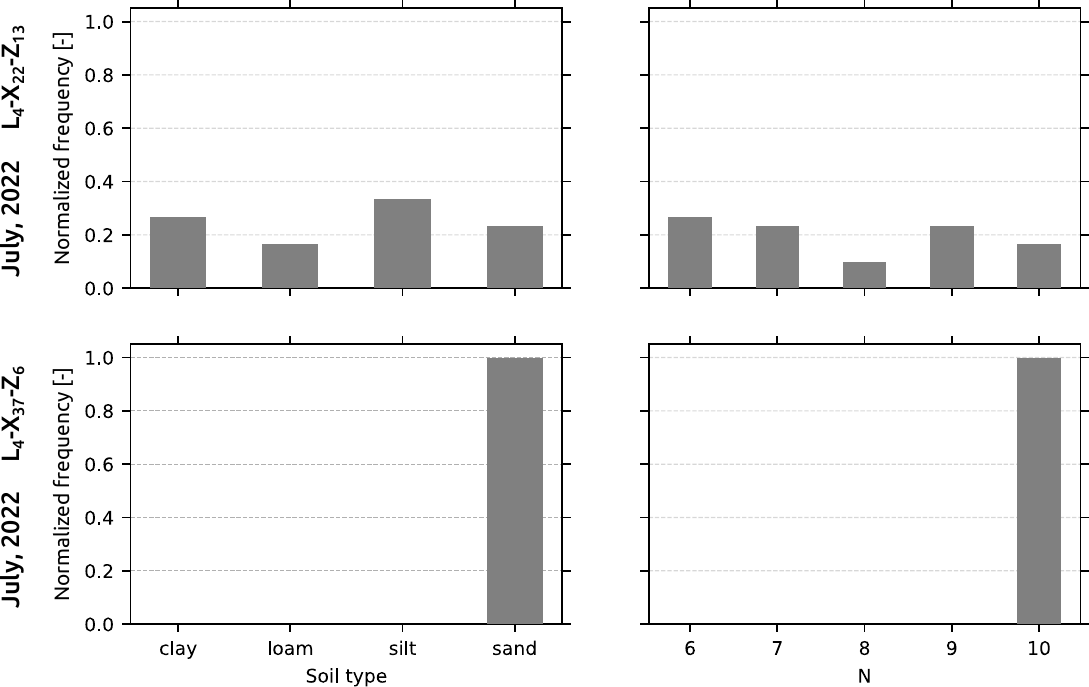}
    \caption{\textbf{Occurrence normalized frequency histograms.}
    Normalized frequencies of the soil type and $N$ value occurrences at two distinct points of the geophone line $L_4$ computed over July, 2022.
    The shown examples correspond to the red annotations in fig.~\ref{fig:FigureS17}.
    }
    \label{fig:FigureS16}
\end{figure}

Although \textsc{Silex} deterministic approach does not directly provide a measure of uncertainty for each prediction, we calculate the inference variability based on the daily results over July 2022 and 2023 (Fig.~3 of the main text, or fig.~\ref{fig:FigureS15}).
With the hypothesis that the petrophysical parameters (soil type, $N$, and at a lesser degree the WT level) should remain stable over a month, we count the number of occurrences of each soil and $N$ at each $(x,y,z)$ point, and divide each number of occurrence by the number of days in the month.
Thus, we obtain the normalized frequency of occurrences of each parameter, and select the soil type and $N$ with the highest frequencies (fig.~\ref{fig:FigureS16}).
High frequency values indicate that the model’s predictions are relatively uniform or concentrated around certain outcomes, suggesting low diversity in the distribution of the inferred petrophysical parameters over the studied month.
In contrast, low frequency values reflect a wider diversity in the daily distribution of outcomes.
We also compute the mean and standard deviation of the WT levels at each $(x,y)$ points.

\begin{figure}[hp]
    \centering
    \includegraphics[width=\textwidth]{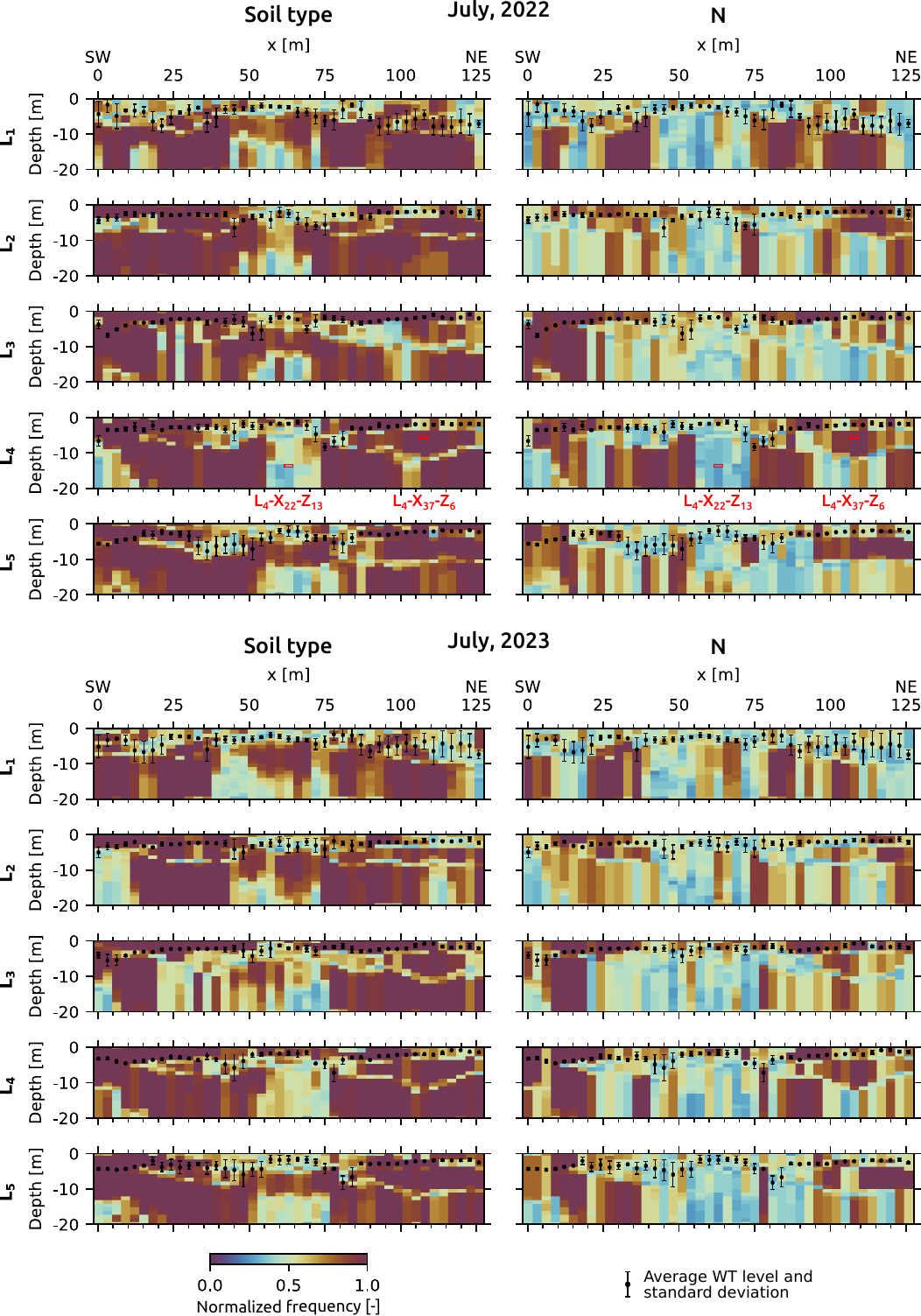}
    \caption{\textbf{Occurrence normalized frequency sections.}
    Normalized frequencies across the five geophone lines ($L_1$ to $L_5$) monthly computed on soil type and $N$ occurrences over July 2022 and 2023, and water table (WT) level with its standard deviation.
    \textbf{Left:} On soil types.
    \textbf{Right:} On $N$ values.
    Red annotations refer to fig.\ref{fig:FigureS16}.
    }
    \label{fig:FigureS17}
\end{figure}

Results for July 2022 and 2023 indicate generally high frequencies across most regions (fig.~\ref{fig:FigureS17}).
However, the frequency is lower (around 0.25) at depth and at the center of all sections, which may be attributed to the lower quality of the input DCs, as discussed in the model evaluation section, that may be more difficult to approximate.
The parameter $N$ shows more variability than the soil type.

It is also worth noting that the spatial boundaries of stiff sand regions are variable over 10~cm (the depth increment), particularly notable on the North-East side of $L_4$ (between $x=42$ and $x=72$~m in figs~\ref{fig:FigureS4} to \ref{fig:FigureS13}).
This suggests that, while these zones are relatively certain, there is still some spatial variability at the very fine scale.

Additionally, geophone line $L_1$ shows higher variability in soil types, $N$ values, and WT levels, in comparison to the other lines, which aligns with the RMSE observed between the input and inferred DCs (figs~\ref{fig:FigureS4} and \ref{fig:FigureS9}).
This suggests that \textsc{Silex} encounters more challenges in converging toward a stable solution.

\clearpage
\newgeometry{left=2.5cm,right=2.5cm,bottom=0.1cm,top=0.1cm} 
\restoregeometry

\subsubsection*{Comparison with conventional seismic inversion}
\label{sec:ComparisonWithconventionalSeismicInversion}
For every geophone line, each DCs along the x-axis (fundamental propagating mode from 15 to 50~Hz), was inverted into $V_S$ over depth logs, to build 2D inverted sections.
We use the open-source software package~\textsc{SWIP}\footnote{\url{https://github.com/spasquet/SWIP}} implemented by~\cite{Pasquet_Bodet_2017}, that is built upon the software~\textsc{Dinver}\footnote{\url{https://www.geopsy.org/wiki/index.php/Dinver:_dinverdc}} using a \textit{neighbourhood algorithm} developed by~\cite{Sambridge_1999} and implemented by~\cite{Wathelet_2008}, to solve the inverse problem in a juxtaposed 1D setup.

This method involves a stochastic exploration of a parameter space in order to search for a minimum \textit{}{misfit} between measured and simulated DCs.
The inversion was parameterized with fours layers and a half-space, in accordance with the drilling data (see $DR_1$ and $DR_2$ in fig.~\ref{fig:FigureS15}, or Fig.~3 of the main text).
The chosen parameter space, as outlined in table~\ref{tab:TableS6}, encompasses various key parameters including layer thicknesses, pressure-wave velocity ($V_P$), shear-wave velocity ($V_S$), density ($\rho$), and Poisson's ratio ($\nu$).
The deliberate selection of a large parameter space stems from the limited \textit{a priori} information about the mechanical properties of the geological layers.
This approach ensures that the inversion process remains explorative, unbiased, and is capable of capturing a wide range of geological scenarios that may influence the seismic response in the study area.

\begin{table}[!htbp]
\centering
  \centering
  \caption{\textbf{Seismic inversion parameter space.}}
  \begin{tabular}{llllll}
      \hline
      Layer $[\#]$ & Thickness $[m]$ & $V_P$ $[m/s]$ & $V_S$ $[m/s]$ & $\rho$ $[kg/m^3]$ & $\nu$ $[-]$\\
      \hline
      1 & 1-3 & 200-600 & 100-300 & 2000-2500 & 0.1-0.5 \\
      2 & 1-3 & 200-1000 & 100-500 & 2000-2500 & 0.1-0.5 \\
      3 & 1-5 & 200-1000 & 100-500 & 2000-2500 & 0.1-0.5 \\
      4 & 1-5 & 200-1500 & 100-750 & 2000-2500 & 0.1-0.5 \\
      half-space & $\infty$ & 600-3000 & 300-1500 & 2000-2500 & 0.1-0.5 \\
      \hline
      \multicolumn{6}{l}{Abbreviations: $V_P$, P-wave velocity; $V_S$, S-wave velocity; $\rho$, density; $\nu$, Poisson's ratio.}
    \end{tabular}
    \label{tab:TableS6}
\end{table}

For each DC along the geophone lines, out of a total 200,400 simulated models, only the models with DCs within the error bars are accepted and averaged to generate final average smooth velocity models.
The DCs' uncertainty $\delta_c(f)$ over frequencies $f$, defining the error-bars, is calculated in accordance with~\cite{ONeill_2003}:
\begin{equation}
     \delta_c(f) = 10^{-a}\left|{\frac{1}{\frac{1}{V_R(f)}-\frac{1}{2fN_x\Delta_x}} - \frac{1}{\frac{1}{V_R(f)}+\frac{1}{2fN_x\Delta_x}}} \right|,
     \label{eq:errorbars}
\end{equation}
with $a$ being the logarithmic reduction factor, usually 0.5, $N_x$ the number of geophones in the MASW window, and $\Delta_x$ the space interval between geophones.
The running parameters used in~\textsc{SWIP} are outlined in table~\ref{tab:TableS7}.

\begin{table}[!htbp]
\centering
  \centering
  \caption{\textbf{Seismic inversion running parameters for~\textsc{SWIP}.}}
  \begin{tabular}{lll}
      \hline
      Parameter & Value & Description \\
      \hline
      $n_{run}$ & 4 & Number of runs\\
      $it_{max}$ & 250 & Number of iterations per run\\
      $ns_0$ & 100 & Number of starting models\\
      $ns$ & 200 & Number of modes created at each iteration\\
      $nr$ & 100 & Number of previous models to build new sub-parameter space\\
      \hline
    \end{tabular}
    \label{tab:TableS7}
\end{table}

\begin{figure}[p]
    \centering
    \includegraphics[width=\textwidth]{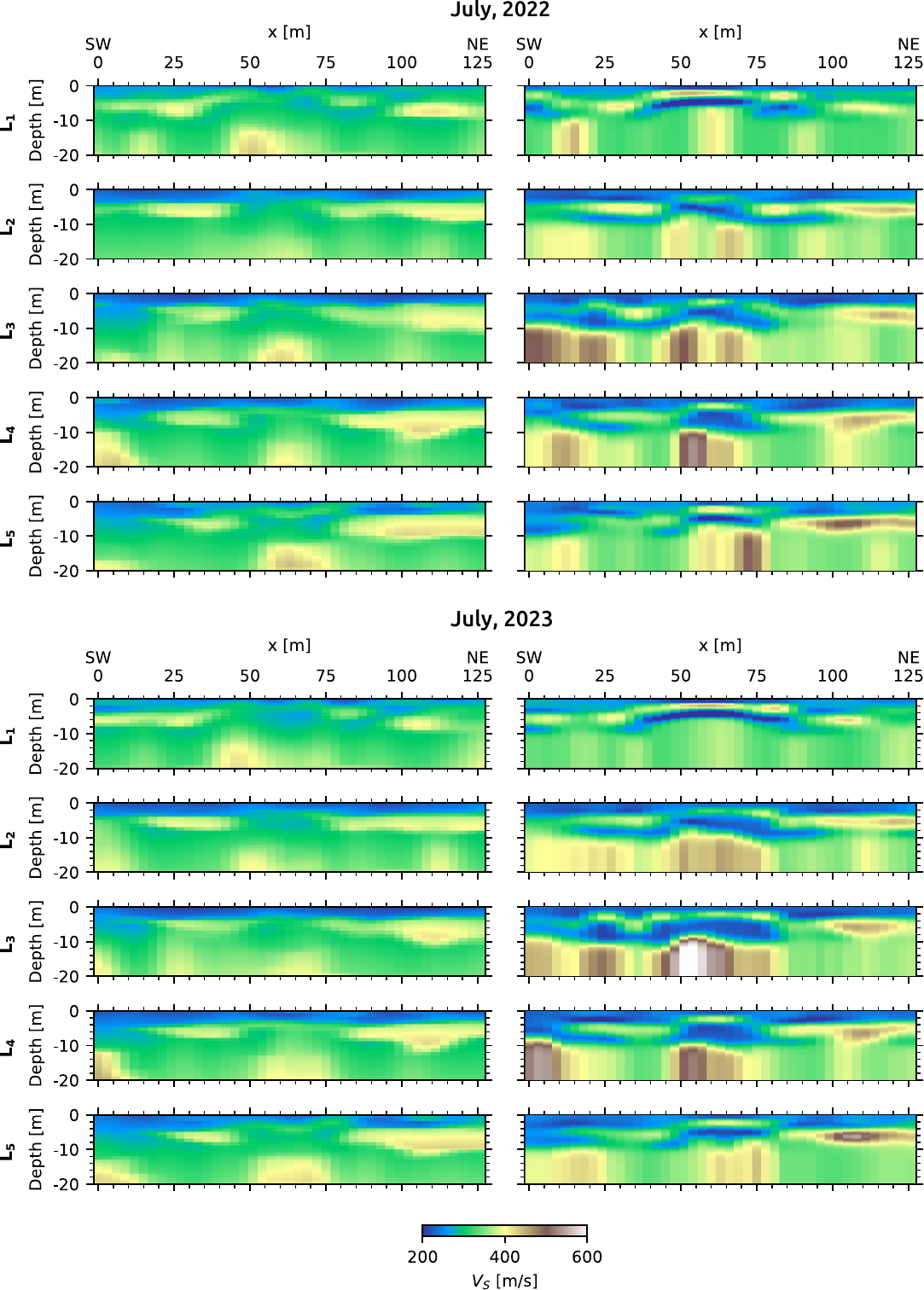}
    \caption{\textbf{Shear-wave velocity sections.}
    Shear-wave velocity ($V_S$) across the five geophone lines ($L_1$ to $L_5$), on July 2022 and 2023.
    \textbf{Left:} Computed using the rock-physics model from \textsc{Silex}'s petrophysical inversion outputs.
    \textbf{Right:} Directly inverted with a conventional seismic inversion method.
    }
    \label{fig:FigureS18}
\end{figure}

The inverted 2D $V_S$ sections for geophone lines $L_1$ to $L_5$, averaged over July 2022 and 2023, obtained through the rock-physics model (based on \textsc{Silex} outputs), are compared to the results from the conventional seismic inversion described above (fig.~\ref{fig:FigureS18}).
Although both methods retrieve similar subsurface structures, we observe some differences in $V_S$ ranges between the two approaches.
This arises because the rock-physics model used to train \textsc{Silex} is limited to soil types and their mechanical properties, only generates smooth $V_S$ models, making it difficult to infer sharp contrasts.
This demonstrates that, while the soil types retrieved by \textsc{Silex} are validated by drillings, its performance is constrained by the limitations of the rock-physics model used to generate the training data.
These limitations are particularly evident in the retrieval of sharp contrasts in mechanical properties, especially for substratum structures composed of rocks rather than soils.

\begin{figure}[p]
    \centering
    \includegraphics[width=\textwidth]{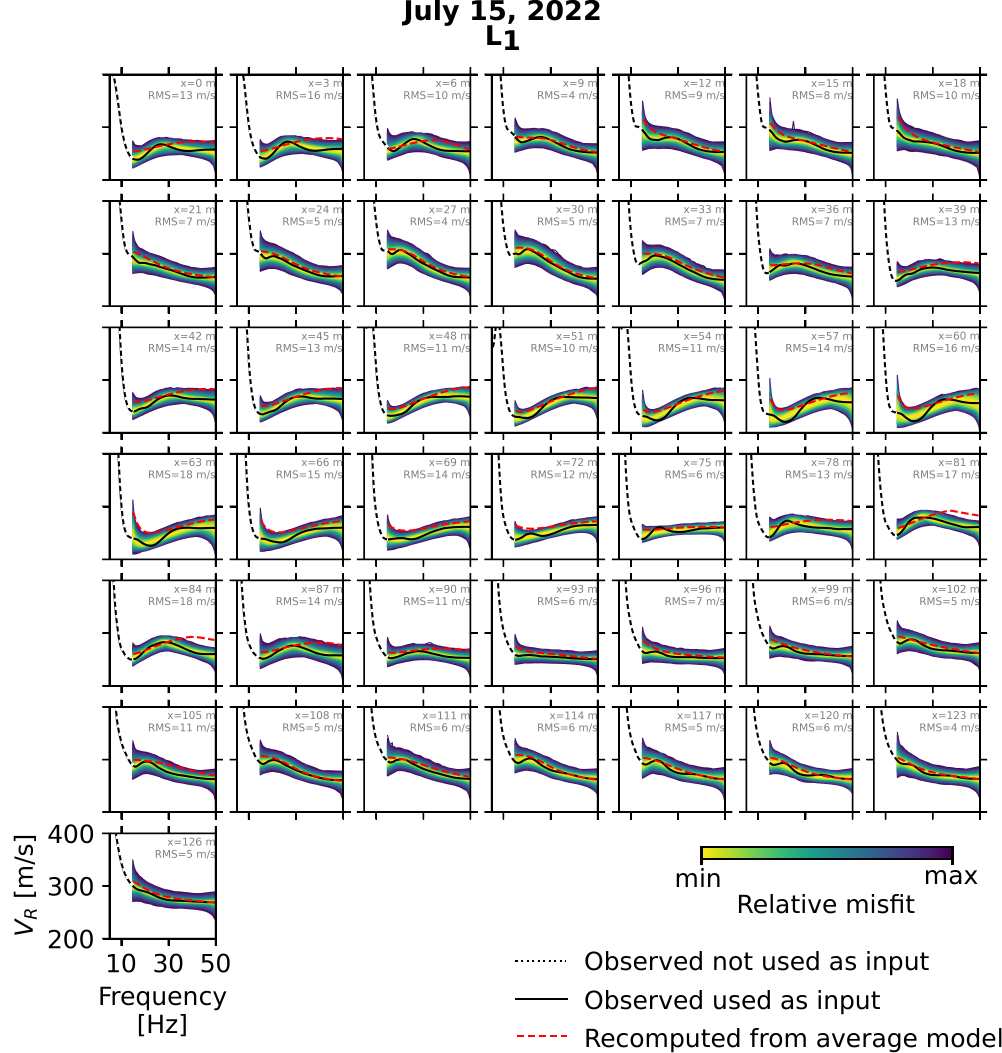}
    \caption{\textbf{Comparison between the real and the conventionally inverted dispersion curves along geophone line $L_1$ on July 15, 2022.} 
    Shown are the input dispersion curves (black) and the recomputed curves from the average models obtained through conventional seismic inversion (red).
    Dispersion curves from all computed models within the error bars (Equation~\ref{eq:errorbars}) are displayed in a color gradient based on their respective misfit values.
    Since the input data only covered frequencies from 15 to 50~Hz, the recomputed curves are similarly limited to this frequency range.
    Positions along the x-axis and root mean square errors (RMSE) are indicated in grey.
    }
    \label{fig:FigureS19}
\end{figure}

\begin{figure}[p]
    \centering
    \includegraphics[width=\textwidth]{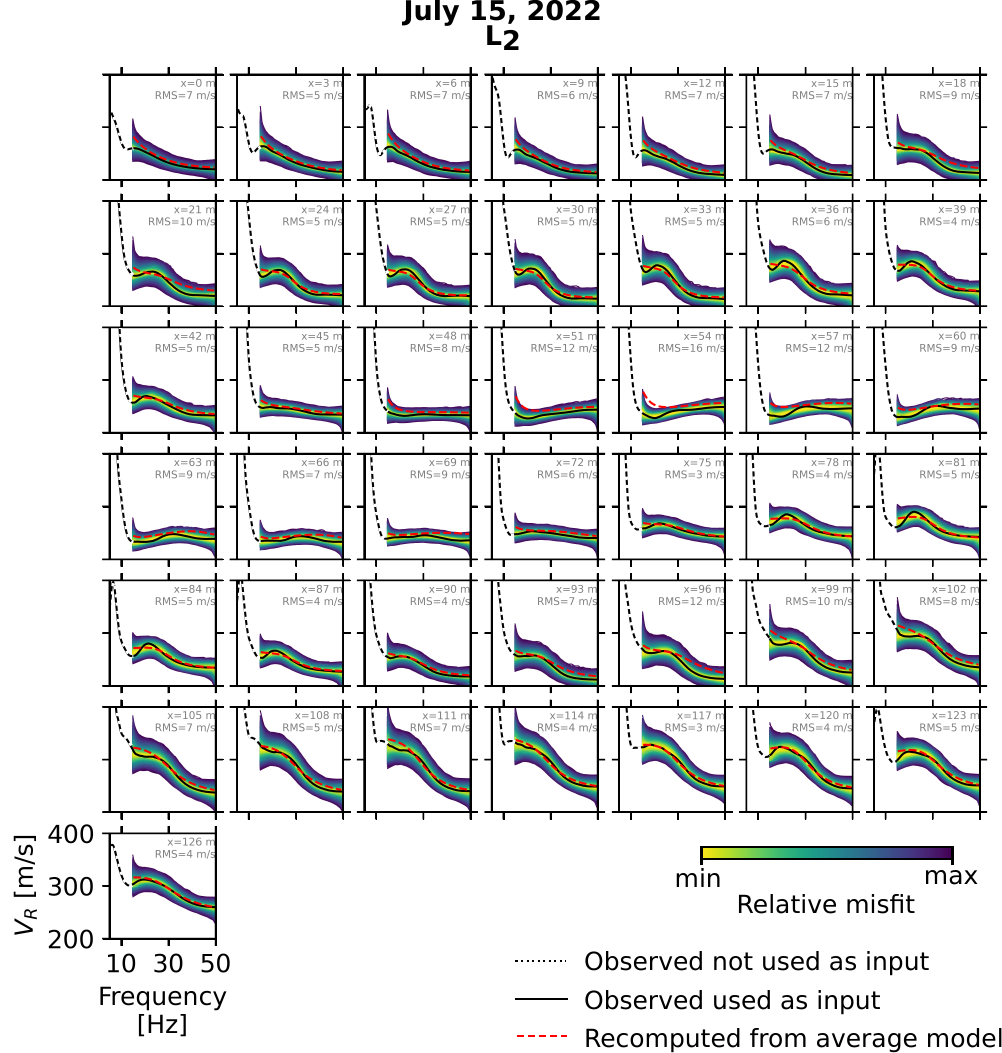}
    \caption{\textbf{Comparison between the real and the conventionally inverted dispersion curves along geophone line $L_2$ on July 15, 2022.} 
    Shown are the input dispersion curves (black) and the recomputed curves from the average models obtained through conventional seismic inversion (red).
    Dispersion curves from all computed models within the error bars (Equation~\ref{eq:errorbars}) are displayed in a color gradient based on their respective misfit values.
    Since the input data only covered frequencies from 15 to 50~Hz, the recomputed curves are similarly limited to this frequency range.
    Positions along the x-axis and root mean square errors (RMSE) are indicated in grey.
    }
    \label{fig:FigureS20}
\end{figure}

\begin{figure}[p]
    \centering
    \includegraphics[width=\textwidth]{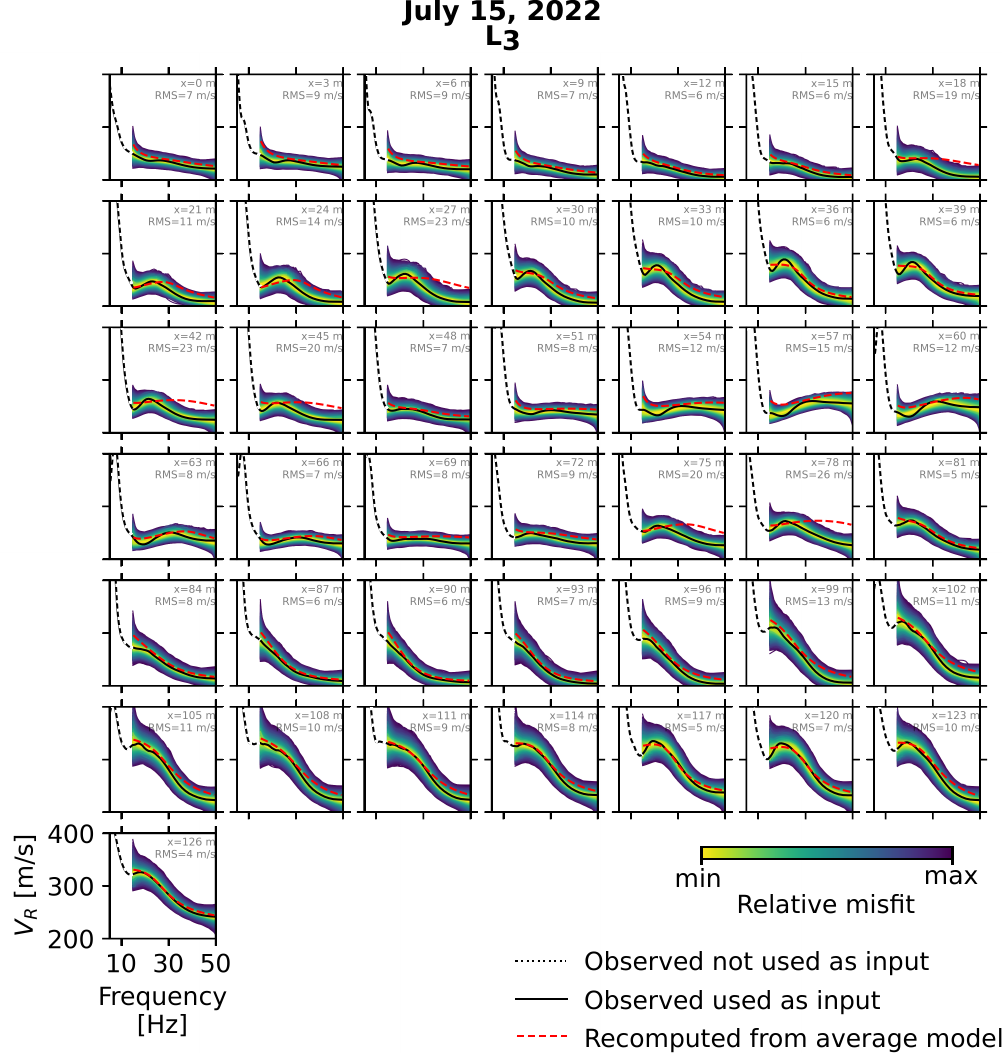}
    \caption{\textbf{Comparison between the real and the conventionally inverted dispersion curves along geophone line $L_3$ on July 15, 2022.} 
    Shown are the input dispersion curves (black) and the recomputed curves from the average models obtained through conventional seismic inversion (red).
    Dispersion curves from all computed models within the error bars (Equation~\ref{eq:errorbars}) are displayed in a color gradient based on their respective misfit values.
    Since the input data only covered frequencies from 15 to 50~Hz, the recomputed curves are similarly limited to this frequency range.
    Positions along the x-axis and root mean square errors (RMSE) are indicated in grey.
    }
    \label{fig:FigureS21}
\end{figure}

\begin{figure}[p]
    \centering
    \includegraphics[width=\textwidth]{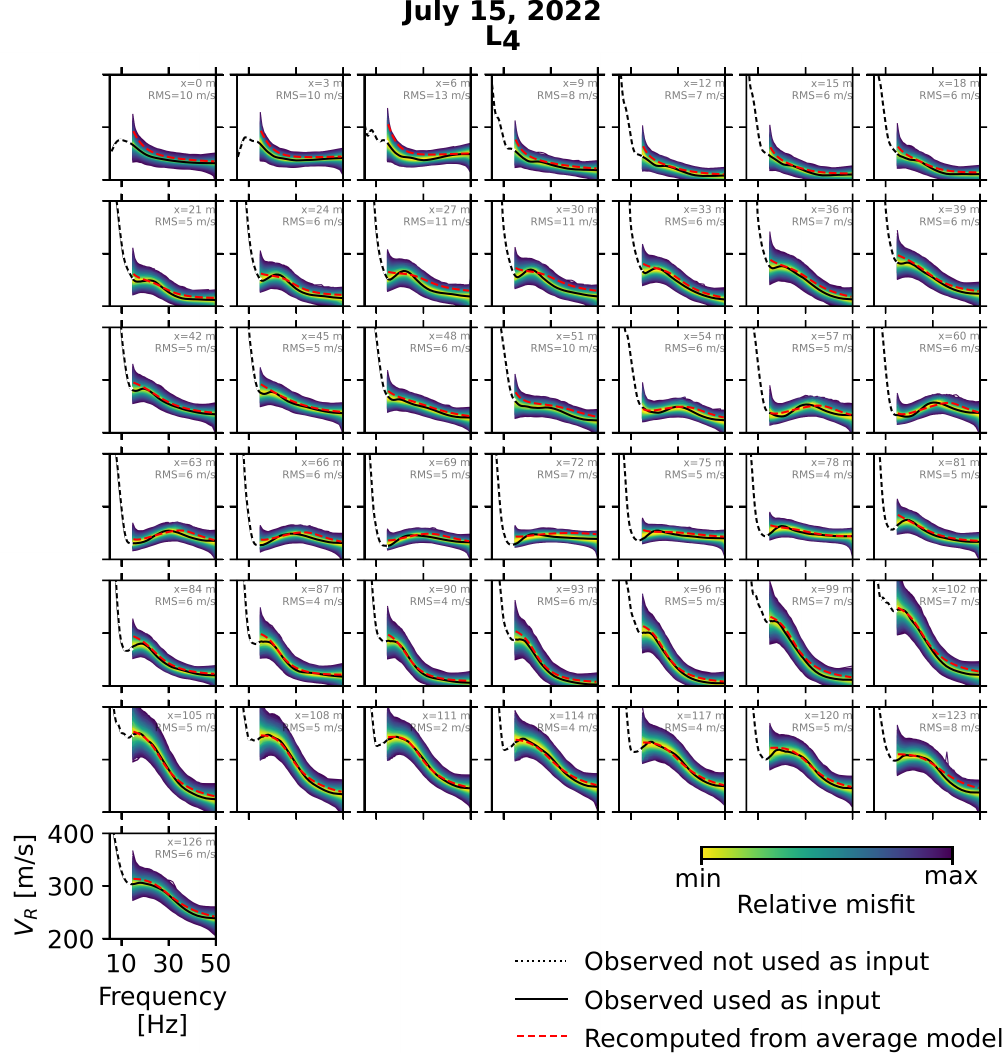}
    \caption{\textbf{Comparison between the real and the conventionally inverted dispersion curves along geophone line $L_4$ on July 15, 2022.} 
    Shown are the input dispersion curves (black) and the recomputed curves from the average models obtained through conventional seismic inversion (red).
    Dispersion curves from all computed models within the error bars (Equation~\ref{eq:errorbars}) are displayed in a color gradient based on their respective misfit values.
    Since the input data only covered frequencies from 15 to 50~Hz, the recomputed curves are similarly limited to this frequency range.
    Positions along the x-axis and root mean square errors (RMSE) are indicated in grey.
    }
    \label{fig:FigureS22}
\end{figure}

\begin{figure}[p]
    \centering
    \includegraphics[width=\textwidth]{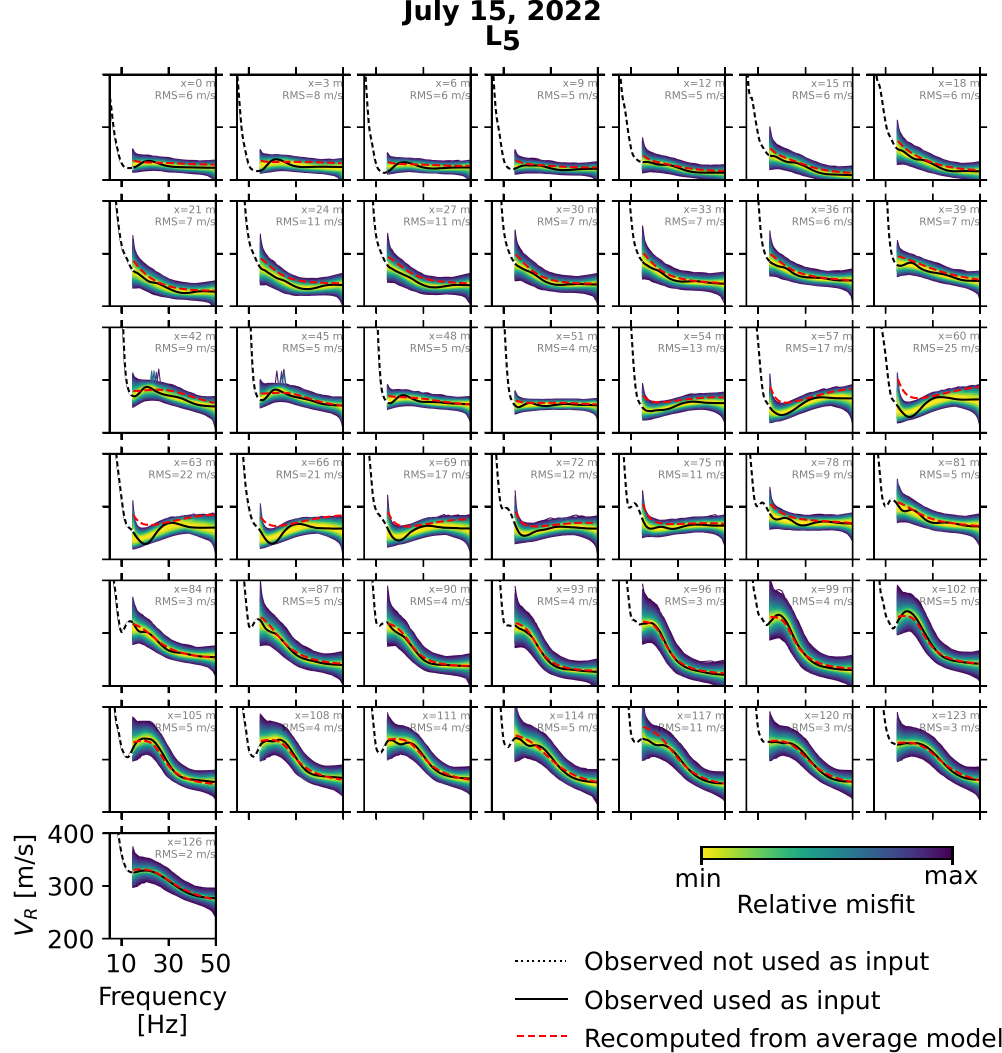}
    \caption{\textbf{Comparison between the real and the conventionally inverted dispersion curves along geophone line $L_5$ on July 15, 2022.} 
    Shown are the input dispersion curves (black) and the recomputed curves from the average models obtained through conventional seismic inversion (red).
    Dispersion curves from all computed models within the error bars (Equation~\ref{eq:errorbars}) are displayed in a color gradient based on their respective misfit values.
    Since the input data only covered frequencies from 15 to 50~Hz, the recomputed curves are similarly limited to this frequency range.
    Positions along the x-axis and root mean square errors (RMSE) are indicated in grey.
    }
    \label{fig:FigureS23}
\end{figure}

\begin{figure}[p]
    \centering
    \includegraphics[width=\textwidth]{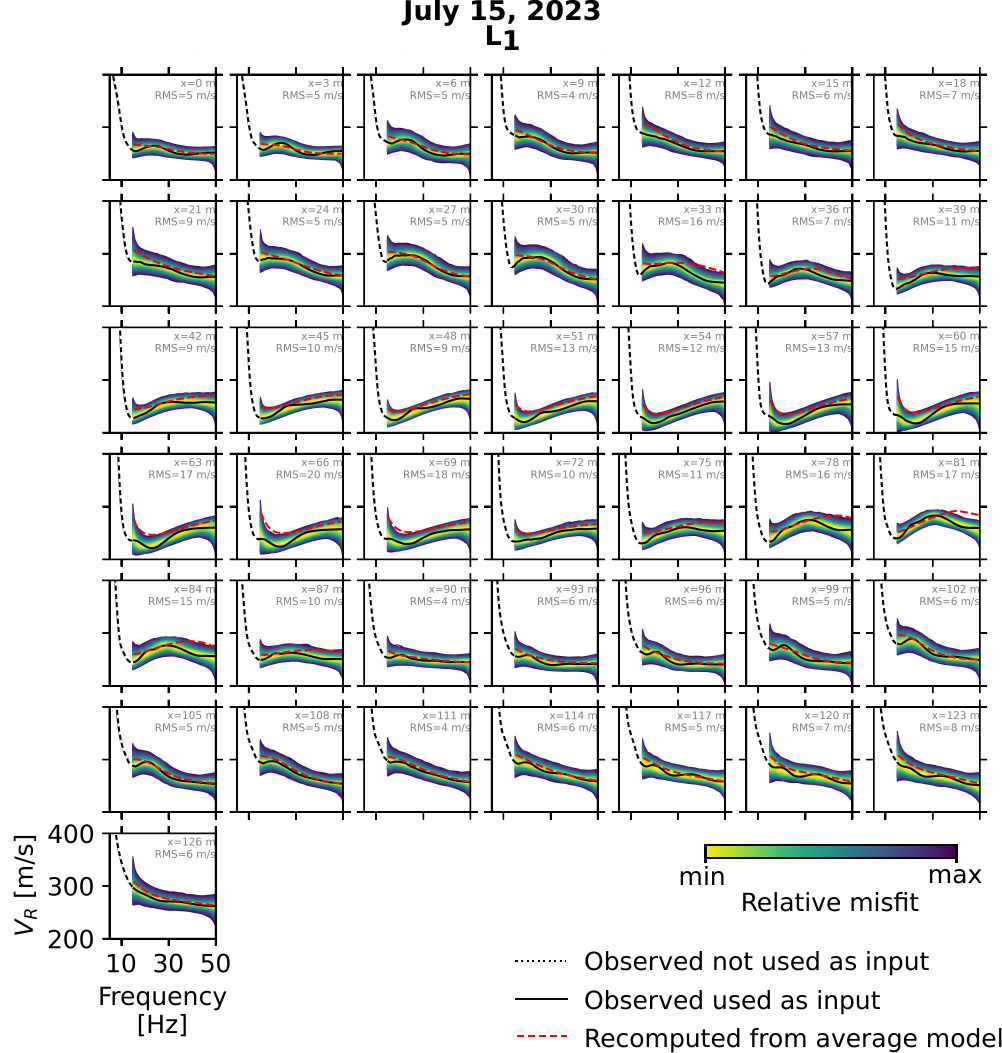}
    \caption{\textbf{Comparison between the real and the conventionally inverted dispersion curves along geophone line $L_1$ on July 15, 2023.} 
    Shown are the input dispersion curves (black) and the recomputed curves from the average models obtained through conventional seismic inversion (red).
    Dispersion curves from all computed models within the error bars (Equation~\ref{eq:errorbars}) are displayed in a color gradient based on their respective misfit values.
    Since the input data only covered frequencies from 15 to 50~Hz, the recomputed curves are similarly limited to this frequency range.
    Positions along the x-axis and root mean square errors (RMSE) are indicated in grey.
    }
    \label{fig:FigureS24}
\end{figure}

\begin{figure}[p]
    \centering
    \includegraphics[width=\textwidth]{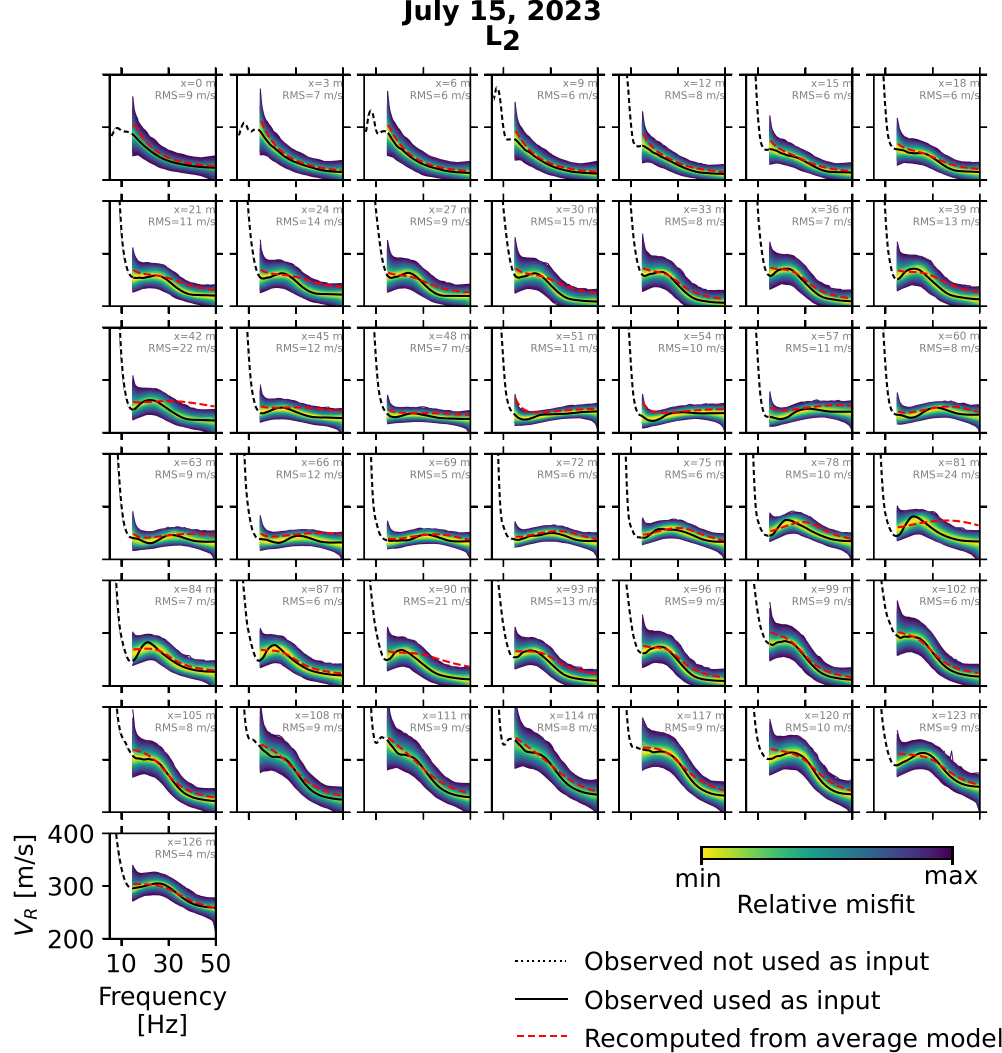}
    \caption{\textbf{Comparison between the real and the conventionally inverted dispersion curves along geophone line $L_2$ on July 15, 2023.} 
    Shown are the input dispersion curves (black) and the recomputed curves from the average models obtained through conventional seismic inversion (red).
    Dispersion curves from all computed models within the error bars (Equation~\ref{eq:errorbars}) are displayed in a color gradient based on their respective misfit values.
    Since the input data only covered frequencies from 15 to 50~Hz, the recomputed curves are similarly limited to this frequency range.
    Positions along the x-axis and root mean square errors (RMSE) are indicated in grey.
    }
    \label{fig:FigureS25}
\end{figure}

\begin{figure}[p]
    \centering
    \includegraphics[width=\textwidth]{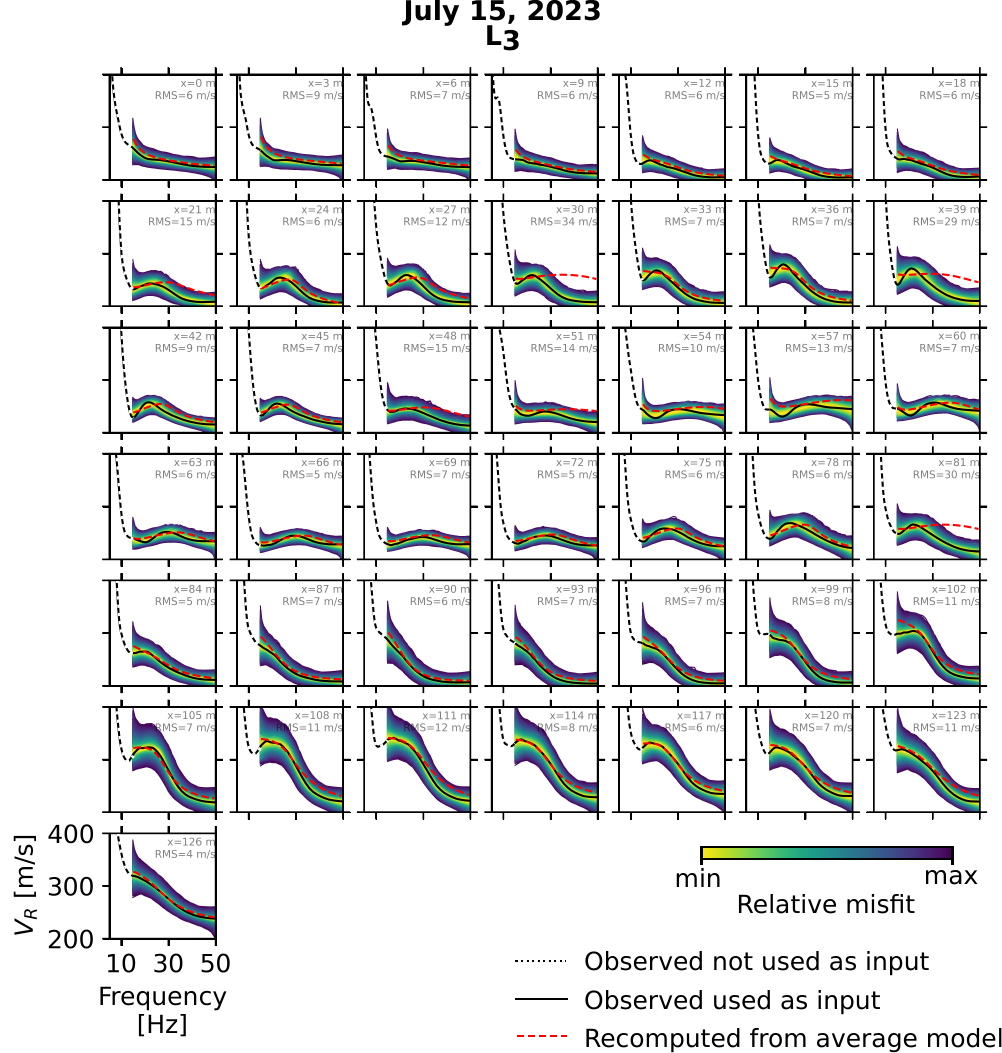}
    \caption{\textbf{Comparison between the real and the conventionally inverted dispersion curves along geophone line $L_3$ on July 15, 2023.} 
    Shown are the input dispersion curves (black) and the recomputed curves from the average models obtained through conventional seismic inversion (red).
    Dispersion curves from all computed models within the error bars (Equation~\ref{eq:errorbars}) are displayed in a color gradient based on their respective misfit values.
    Since the input data only covered frequencies from 15 to 50~Hz, the recomputed curves are similarly limited to this frequency range.
    Positions along the x-axis and root mean square errors (RMSE) are indicated in grey.
    }
    \label{fig:FigureS26}
\end{figure}

\begin{figure}[p]
    \centering
    \includegraphics[width=\textwidth]{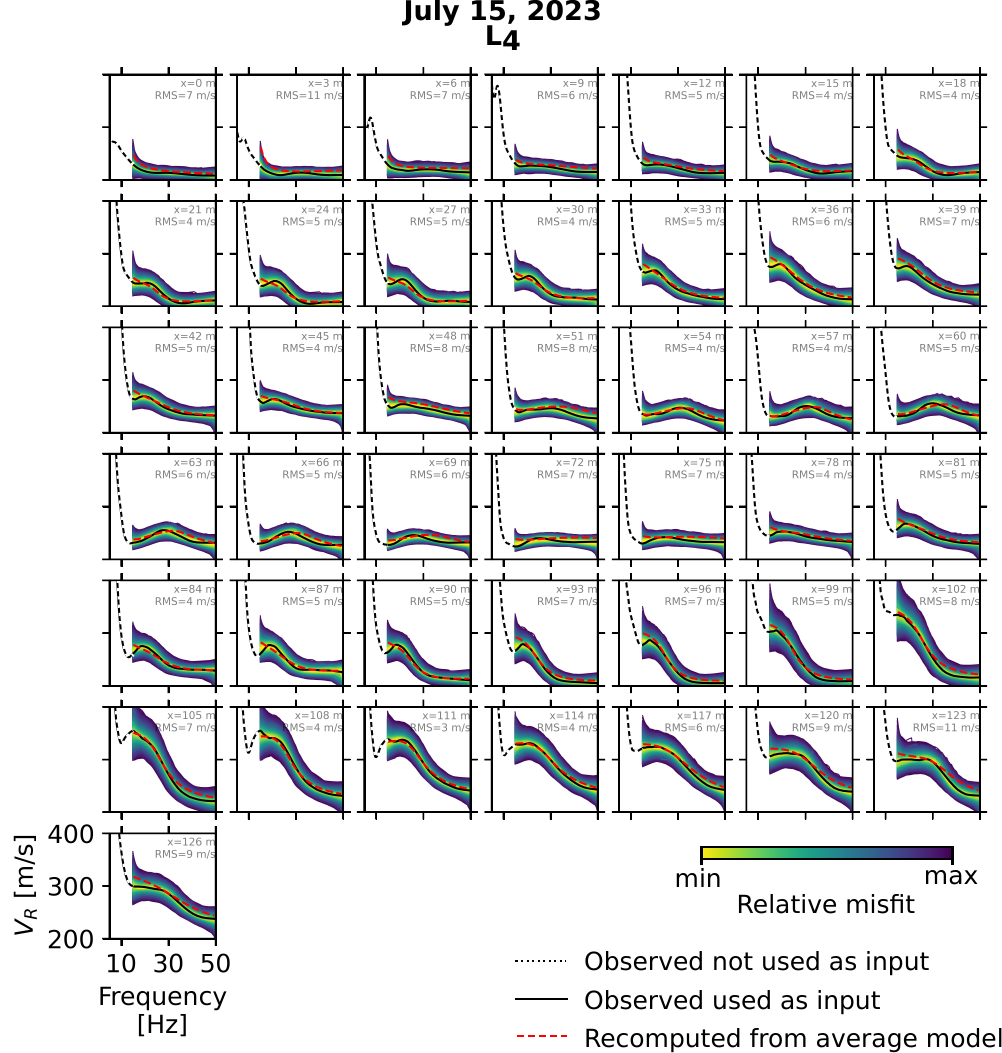}
    \caption{\textbf{Comparison between the real and the conventionally inverted dispersion curves along geophone line $L_4$ on July 15, 2023.} 
    Shown are the input dispersion curves (black) and the recomputed curves from the average models obtained through conventional seismic inversion (red).
    Dispersion curves from all computed models within the error bars (Equation~\ref{eq:errorbars}) are displayed in a color gradient based on their respective misfit values.
    Since the input data only covered frequencies from 15 to 50~Hz, the recomputed curves are similarly limited to this frequency range.
    Positions along the x-axis and root mean square errors (RMSE) are indicated in grey.
    }
    \label{fig:FigureS27}
\end{figure}

\begin{figure}[p]
    \centering
    \includegraphics[width=\textwidth]{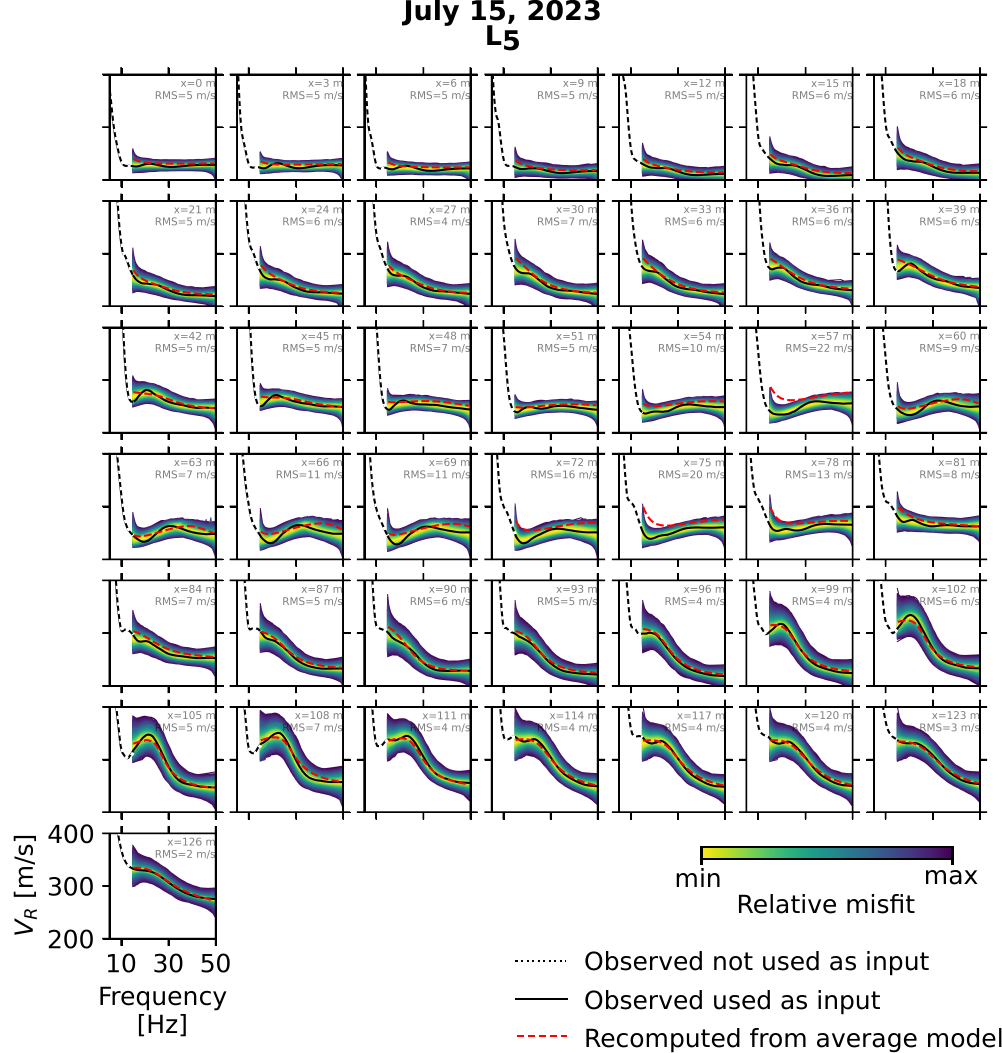}
    \caption{\textbf{Comparison between the real and the conventionally inverted dispersion curves along geophone line $L_5$ on July 15, 2023.} 
    Shown are the input dispersion curves (black) and the recomputed curves from the average models obtained through conventional seismic inversion (red).
    Dispersion curves from all computed models within the error bars (Equation~\ref{eq:errorbars}) are displayed in a color gradient based on their respective misfit values.
    Since the input data only covered frequencies from 15 to 50~Hz, the recomputed curves are similarly limited to this frequency range.
    Positions along the x-axis and root mean square errors (RMSE) are indicated in grey.
    }
    \label{fig:FigureS28}
\end{figure}

As for figs~\ref{fig:FigureS4} to \ref{fig:FigureS13}, figs~\ref{fig:FigureS19} to \ref{fig:FigureS28} show the comparison between the input and recomputed DCs, from the conventional inversion results, on July 15, 2022 and 2023, over the 5 geophone lines.
Each DC is represented with a color depending on the misfit value ($MF$) between the experimental data and the simulated dispersion defined as: 
\begin{equation}
    MF = \sqrt{\sum_{i=1}^{N_f}{\frac{(V_{{sim}_i}-V_{{exp}_i})^2}{N_f\delta_{ci}^2}}}
    \ ,
\end{equation}
with $V_{{sim}_i}$ and $V_{{exp}_i}$ being the simulated and experimental phase velocities at each frequency $f_i$, $N_f$ the number of frequency samples, and $\delta_{ci}$ the phase-velocity measurement uncertainty (error bars) at each frequency $f_i$.
The average root mean square error (RMSE, see Equation~\ref{eq:RMSE}) on the DCs was estimated at 9~m/s (against 12~m/s for \textsc{Silex} inferences).

\clearpage
\newgeometry{left=2.5cm,right=2.5cm,bottom=0.1cm,top=0.1cm} 
\restoregeometry

\clearpage

\clearpage

\clearpage

\clearpage

\clearpage

\clearpage

\clearpage

\clearpage

\clearpage

\clearpage



\end{document}